\documentclass{article}
\usepackage{hyperref}
\usepackage{morefloats}
\RequirePackage[OT1]{fontenc}
\RequirePackage{amsthm,amsmath}
\RequirePackage{natbib}
\usepackage[margin=1.17in]{geometry}
\usepackage{graphicx}
\usepackage{amssymb}
\usepackage{etoolbox}
\patchcmd{\thebibliography}{\chapter*}{\section*}{}{}
\usepackage{empheq}
\usepackage{keyval}
\usepackage{caption}
\usepackage{subcaption}
\usepackage{titling}
\usepackage{pdfpages}
\usepackage{tikz}
\usepackage{tikz,tkz-base}
\usepackage{mdframed}
\usepackage{tabularx}
\usepackage{array}
\usepackage{siunitx}
\usepackage{nicefrac}
\usepackage{xr}
\usetikzlibrary{shapes,decorations,shadows}
\usetikzlibrary{decorations.pathmorphing}
\usetikzlibrary{decorations.shapes}
\usetikzlibrary{fadings}
\usetikzlibrary{patterns}
\usetikzlibrary{calc}
\usetikzlibrary{decorations.text}
\usetikzlibrary{decorations.footprints}
\usetikzlibrary{decorations.fractals}
\usetikzlibrary{shapes.gates.logic.IEC}
\usetikzlibrary{shapes.gates.logic.US}
\usetikzlibrary{fit,chains}
\usetikzlibrary{positioning}
\usepgflibrary{shapes}
\usetikzlibrary{scopes}
\usetikzlibrary{arrows}
\usetikzlibrary{backgrounds}

\tikzset{latent/.style={circle,fill=white,draw=black,inner sep=1pt, 
minimum size=20pt, font=\fontsize{10}{10}\selectfont},
obs/.style={latent,fill=gray!25},
const/.style={rectangle, inner sep=0pt},
factor/.style={rectangle, fill=black,minimum size=5pt, inner sep=0pt},
>={triangle 45}}

\pgfdeclarelayer{b}
\pgfdeclarelayer{f}
\pgfsetlayers{b,main,f}

\numberwithin{equation}{section}
\theoremstyle{plain}

\usepackage[space]{grffile}
\usepackage{enumitem}

\newcommand{\getFullPath}[1]{#1}

\begin{document}

\setlength{\abovedisplayskip}{5pt}
\setlength{\belowdisplayskip}{5pt}

\title{Modeling Recovery Curves With Application to Prostatectomy}






\author{Fulton Wang$^\diamond$, Cynthia Rudin$^\bullet$, Tyler
  H. Mccormick$^\clubsuit$, John L. Gore$^\triangle$\\[4pt]
\textit{$^\diamond$EECS Department, Massachusetts Institute of Technology}
\\[2pt]
\textit{$^\bullet$Computer Science Department, Duke University}
\\[2pt]
\textit{$^\clubsuit$Statistics Department and Sociology Department, University of Washington}
\\[2pt]
\textit{$^\triangle$Department of Urology, University of Washington}
\\[2pt]
{fultonw@mit.edu}}

\markboth%
{F. Wang and others}
{Modeling Recovery Curves}

\maketitle

\begin{abstract}
{In many clinical settings, a patient outcome takes the form of a scalar time series with a recovery curve shape, which is
characterized by a sharp drop due to a disruptive event (e.g.,
surgery) and subsequent monotonic smooth rise towards an asymptotic
level not exceeding the pre-event value.  We propose a Bayesian model that predicts recovery curves based on
information available before the disruptive event.  A recovery curve
of interest is the quantified sexual function of prostate cancer
patients after prostatectomy surgery. We illustrate the utility of our
model as a pre-treatment medical decision aid, producing personalized
predictions that are both interpretable and accurate.  We uncover covariate relationships that agree with and supplement that in existing medical literature.}
\end{abstract}

\section{Introduction\label{sec:intro}}
In the medical community, there is a pressing need for personalized
predictions of how a disruptive event, such as a treatment or disease,
will impact particular bodily function levels. Of particular interest
is the extent to which the function is initially perturbed by the
event and the ensuing pattern of recovery. In many contexts, such as
mental acuity following a stroke or sexual function following
prostatectomy, the post-event trajectory generally exhibits what we call a recovery curve shape, characterized by an initial instantaneous drop followed by a monotonic rise towards an asymptotic level not exceeding the original function level. Here, we propose a Bayesian model that can be used to predict a patient's expected recovery curve, given information about the patient that is available before the event.

This paper presents a decision aid for patients considering a
medical treatment who want to know what adverse side effect the
treatment would have on a particular bodily function.  In particular,
our model will be used to display to the patient a distribution over
post-treatment function trajectories, conveying the uncertainty in
predictions that should be considered in decision-making.  We assume
that the function level lies in some closed interval, the
pre-treatment function level is known, and the adverse effect of the
event on the function is a priori known by the medical community to be immediate, but wearing off over time.

If a model is to be widely adopted as a medical decision aid, it is
not enough for it to merely produce predictions that are accurate; it
must also be \emph{interpretable}: it must give predictions that a healthcare provider or patient can
readily understand.  This is crucial in a clinical setting not only
because of time constraints, but also because each additional point of
confusion regarding the predictions decreases the flow of information
to the patient and thus their trust in it.  Any model, including ours,
should be used only if it fits the data, and standard model checking
should be performed.  However, aggregate model fit and predictive accuracy are
not sufficient; when a tool causes more
questions than it answers, it will be less likely to be used.

In the applications we consider, we
will predict time series that are expected to be recovery curves.  These recovery curves occur in practice in situations where a medical procedure may cause a temporary inhibition or detriment in a patient, but will not improve the patient's condition over a baseline value.   In our case, we examine sexual function in patients with prostate cancer who are considering a prostatectomy.  Though a prostatectomy will likely decrease sexual function in some patients, it will not \emph{improve} a patient's sexual function above the baseline.  

We restrict the
space of possible outputs to our model, therefore, to \emph{only}
include predictions that are recovery curves.  In other words, our model outputs
a distribution of post-event trajectories each of which is guaranteed
to be a recovery curve, so that the function level drops
instantaneously downwards at event time, and rises smoothly to
approach an asymptotic level lying between the pre-treatment value and
value immediately following treatment (we will use the words ``level''
and ``value'' interchangeably).  Furthermore, our model encourages the
posterior predictive distribution over trajectories to have a well
defined mode, so that the distribution of curves, when plotted, can be
visually easily interpreted as a single maximum a posteriori prediction
along with the uncertainty in that single prediction.  Thus, not only
do the predictions match prior expections, but they are also simple,
defined by a small set of salient features (i.e. initial drop,
recovery rate, and range of uncertainty), not
containing extraneous artifacts that slightly increase accuracy but greatly reduce interpretability.

We use our method to create a decision aid for prostate cancer
patients who are considering a prostatectomy, predicting their sexual
function trajectory, should they undergo a prostatectomy.  We fit our model
using data from a study that tracked the quantified
sexual function level, expressed as a number between 0 and 1, of 237 patients both before radical prostatectomy surgery and at
a common set of timepoints in the 4 years immediately following
surgery. These numerical measures of sexual function were obtained by
administering the Prostate Cancer Index, which is a multiple choice
questionaire that first evaluates patient function and
bother following prostate cancer treatment, and then converts the
answers to a numerical score.  Note that while our target population
is patients merely considering a prostatectomy (and satisfies two
properties listed shortly), our dataset only contains data from
patients who actually did undergo a prostatectomy.

Prostate cancer will affect 1 out of 6
men, and low stage patients usually have several viable treatment
options, each with different side effects. Radical prostatectomy is known to
adversely affect sexual function.  Thus, for patients considering prostatectomy, it is important to
forecast the pattern of sexual function level should they undergo one.  Past studies \citep{Potosky2004} and our dataset
indicate that sexual function trajectories after
prostatectomy follow a recovery curve (at least up to 5
years post-treatment), suggesting our model may fit such trajectories well.

\begin{figure}[]
\centering
\begin{subfigure}{0.48\linewidth}
\includegraphics[width=1\linewidth]{\getFullPath{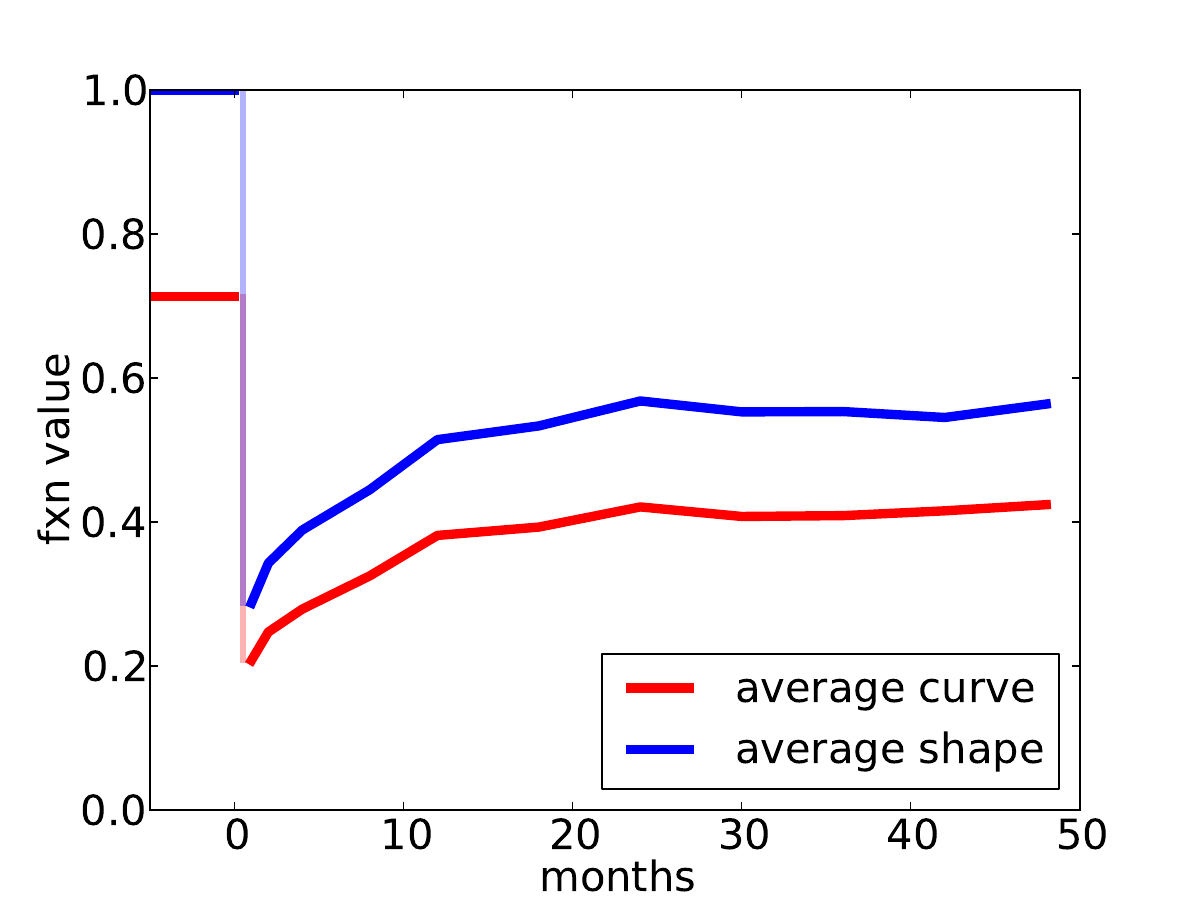}}
\caption{The average function value and scaled function value over the prostatectomy dataset exhibit a recovery curve shape.}
\label{fig:avg_curve}
\end{subfigure}
\quad
\begin{subfigure}{0.48\linewidth}
\includegraphics[width=1\linewidth]{\getFullPath{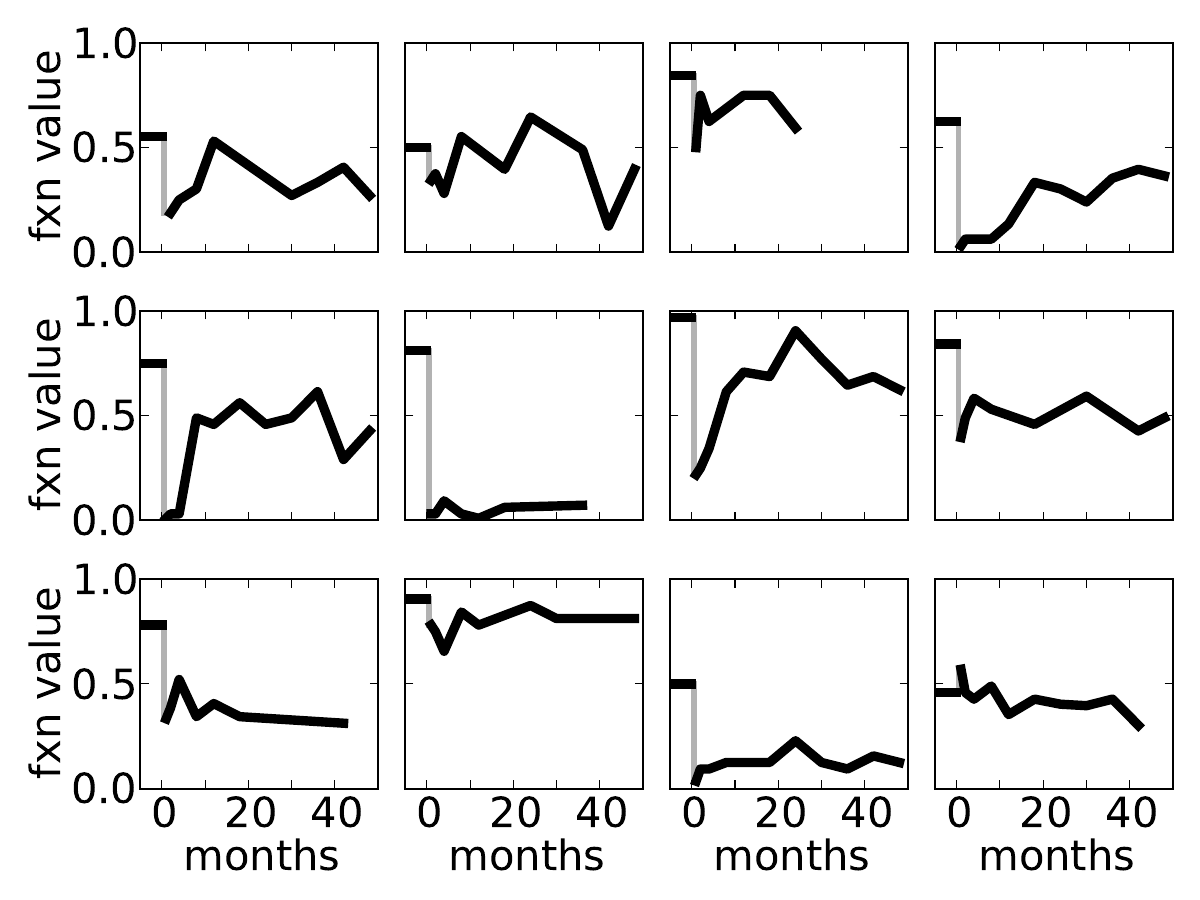}}
\caption{Raw data of 12 randomly chosen patients who passed the filters as described in Section \ref{sec:data}.\\}
\label{fig:example_curves}
\end{subfigure}
\caption{Sexual function trajectory (\texttt{fxn value}) following prostatectomy.}
\label{fig:data_plots}
\end{figure}

To illustrate, in Figure
\ref{fig:avg_curve} we show the dataset-wide averages, for each
timepoint, of sexual function level (\texttt{fxn level}), as well as dataset-wide averages,
for each timepoint, of the patients' sexual function values scaled by
their respective value immediately before prostatectomy. We also show
in Figure \ref{fig:example_curves} the unscaled sexual function level time
series of 12 randomly selected patients, which includes their unscaled
sexual function
level both post-treatment and right before treatment.  We hypothesize that the
function levels reported by individual patients are noisy versions of
a latent smooth ``true'' function level.  In this context, we use our method to study whether there are patient covariates that correlate with post-surgery sexual function trajectory.

 The remainder of the paper is organized as follows.  In
 Section~\ref{sec:past} we contextualize our approach by describing
 related work.  Then, we formally define the recovery curve shape in
 Section~\ref{sec:recovery} and present the specifics of our model in
 Section~\ref{sec:model}.  We demonstrate that the model performs well
 in simulation studies in Section~\ref{sec:sim} and then using the
 aforementioned prostatecomy data 
 in Section~\ref{sec:real}.  

\section{Related Work\label{sec:past}}
Past work on personalized prediction of sexual function following prostate cancer treatment has attempted to predict a post-operative binary outcome, typically whether one is able to achieve an erection sufficient for sex at some single timepoint following treatment \citep{Regan2011,Descazeaud2006,Eastham2008, Ayyathurai2008}, or the change in \citep{Vordermark2008} or absolute level \citep{Talcott2003} of some continuous measure of sexual function such as the IIEF-5 score.  Such models incorporated patient covariates in  linear regression models for continuous outcomes, and logistic regression models for binary outcomes even though sexual function is not a binary outcome \citep{Briganti2011}.  Another deficiency of logistic and linear regression is that they are not suitable for modeling longitudinal outcomes, whereas a patient would want to know their entire post-treatment function trajectory.  The only longitudinal model in the literature uses linear regression to relate the change in function level between two fixed timepoints post-treatment \citep{Potosky2004}.

Functional data analysis \citep{Ramsay2002,Denison1998} and growth
curve modeling \citep{Jung2008} are rich areas of past study, but
existing models from those fields do not guarantee that the predicted
time series possesses a recovery curve shape, that it drops following the event and then monotonically approaches
an asymptotic level no higher than the pre-event value.  Parametric
functions mentioned in \citet{Rogosa1985} resemble the functional form
we assume of recovery curves. However, they are modeling growth, not
recovery after some disruptive event, and assume the initial level of the growth curve to be
known, whereas we are trying to predict the entire post-treatment
trajectory, which includes the initial post-treatment value.
Furthermore, they do not place an upper bound on the asymptotic level
of the predicted function.  Isotonic regression models
\citep{Shively2009, Mammen, Neelon2004, Cai2007}, enforce the
predicted functions to be monotonic, but do not naturally output
recovery curves as predictions.

Other statistical models have also been applied
in contexts where the predicted time series is expected to exhibit a
recovery curve shape.  For example, in a medical context, growth curve
techniques have been used to model recovery of a bodily function
following a disruptive event.  \citet{Warschausky2001} model recovery
of FIM-measured function following spinal surgery as being the sum of
a linear and a plateauing function.  Although the FIM-score must lie
within a bounded interval, they do not guarantee that the predicted scores lie within that interval.  \citet{Rolfe2011} model verbal function following chemotherapy using a Bayesian latent basis model, but the model lacks incorporation of patient-correlates, and is instead specialized to infer average recovery at only two fixed timepoints. \citet{Tilling2001a} model a measure of quality of life - the Barthel Index - following stroke using a multilevel model where both patient-specific and time-specific contributions are modeled as a linear combination of a fractional polynomial basis.  In all these models, the predicted time series is expected to possess a recovery curve shape, but there is no explicit constraint built into the models to ensure that the predicted trajectory actually does possesses a recovery curve shape.

Two past works suggest that a model of sexual function following
prostatectomy needs a significant amount of interpretability-promoting
features if it is to be used in practice.  One work involved eliciting and incorporating the preferences of
patients, providers, and design experts via a 3-step human-centered
design process to design such dashboards \citep{hartzler2015design}.  The second work measured patients' abilities to understand 3 different
visual displays communicating the same information - bar charts, line
graph, and table, and examined how a patient's understanding of
those 3 displays related to their demographics
(i.e. education level, race) and graphical and numerical literacy as
measured through the REALM and SNS questionaires, two standard medical
instruments \citep{nayak2015relevance}.  One takeway from these works is just how much detail, care, and user
feedback goes into designing visual dashboards and studying the impact
of seemingly small changes to them.  One example of a small design
decision significant enough to warrant study was, in a pictogram used
to communicate well-being, whether a sunny weather/cloudy weather icon
should be used in place of a smiling/frowning face to represent well being.  These meticulous studies reflect the
sensitivity of patient comprehension to dashboard features, and suggest each
additional feature improving the interpretability of our model can
greatly improve patient comprehension and thus clinical applicability.  A second takeaway from these
works is that some patients prefer extremely simple dashboards. For
example, some patients felt that putting confidence intervals on
personalized predictions was too confusing, preferring 
a point prediction instead.  This preference bolsters the case for our unimodality requirement -
given that some patients do not even want to see uncertainty in
predictions, were we to display them, we should do so in the
simplest manner possible.  In any case, this second takeaway
suggests many interpretability-promoting features may be necessary for
any clinical applicability at all.
\vspace{-0pt}
\section{Recovery Curves\label{sec:recovery}}


\subsection{Recovery Curve Definition\label{subsec:recovery_def}}
A \emph{recovery curve} is a function $f(t)$ defined on $\mathbb{R^+}$.  We will always define $f(t)$ piecewise as
\begin{equation}
 f(t) =
  \begin{cases} 
      \hfill S    \hfill & \text{ for $t=0$}, \\
      \hfill f^+(t) \hfill & \text{ for $t > 0$,} \\
  \end{cases}
\end{equation}which is interpreted as the disruptive event occuring right after time 0, so that $S$ is the (known) pre-treatment function value, and $f^+(t)$ is the post-treatment trajectory.  A recovery curve will satisfy the following:  
\begin{align}
(f^+)' &> 0 \text{ for  t $>$ 0}, \label{req:1}\\
f^+(t) &\leq S \text{ for  t $>$ 0, \label{req:2}}\\
f^+(t) &\geq 0 \text{ for  t $>$ 0, \label{req:3}}\\
S &\in [0,1] \label{req:4}.
\end{align}

\subsection{Parameterizing Recovery Curves\label{subsec:recovery_param}}

We parameterize $f(t)$ scaled to the pre-treatment function value, instead of $f(t)$ itself, assuming:
\begin{equation}f^+(t;S,\theta) = S g(t;\theta). \end{equation}

Thus, the actual post-event trajectory is the shape of the post-event
trajectory, $g(t;\theta)$, scaled to the pre-event function level.  We
choose this parameterization because in order to satisfy requirements of
Equations \ref{req:2} and \ref{req:3}, we just need to ensure that
$g(t;\theta) \in (0,1)$ for $t>0$.  In this work we will refer to the
scaled post-event trajectory, denoted $g(t;\theta)$, as a
\emph{recovery shape}, and use the term \emph{scaled function
  value} to refer to a patient's function value normalized by their
pre-treatment value; a patient's recovery shape is a
time series over their scaled function values.

We parameterize recovery shape $g(t;\theta)$ with 3 parameters: $A$,  the asymptotic drop in scaled function level after surgery,  $B$, the initial drop in scaled function value in excess of the asymptotic drop, and $C$, the rate of recovery of the scaled function value.

\vspace{-0pt}
\begin{align}
  f(t;S,A,B,C) &= Sg(t;A,B,C) \label{eq:shape},\ \text{where}\\
g(t;A,B,C)  &=S\left(1-A - B(1-A)\exp\left(^-\frac{t}{C}\right)\right), \label{eq:f}\\
  A \in [0,1] ,
  B \in [0,1] ,
  C \geq 0 \label{req:7}.
\end{align} 

$f(t;S,A,B,C)$ is a recovery curve if the constraints in Equations \ref{req:1} to \ref{req:4} are satisfied, and we respect those constraints in our model.

\section{Model\label{sec:model}}

The previous section formalized the definition or a recovery curve.  To fit the curves to data and make meaningful predictions for patients, however, we need both the shape of the curve and a framework for inference.  In this section, we describe a Bayesian approach to fitting recovery curves.  We first describe the structure of the statistical model, then we note the properties of the model that make it well-suited for situations where we expect recovery curve trajectories. 

Throughout the section, we refer
to the $i\text{-}th$ patient's covariate vector as $X_i \in
\mathbb{R}^K$ where $K$ is the number of features per patient, and
their observed function
value at time $t$ by $y_i (t)$.  Here, $y_i (t)$ is considered a noisy measurement of
their ``underlying'' function value at time $t$,
$f(t;S_i,A_i,B_i,C_i)$, where
$f(\cdot;S_i,A_i,B_i,C_i)$ is the parameterization of
post-treatment function value described in Section
\ref{subsec:recovery_param}.  The noise in $y_i(t)$ could arise from a number of sources, including short-term fluctuations in patients' experiences or difficulty in recalling function between time periods.  The ``underlying'' function value, $f(t;S_i, A_i,B_i,C_i)$, is a
function of the patient's pre-treatment function value $S_i$ and patient-specific random parameters
$A_i,B_i,C_i$.  To simplify notation, we abbreviate the latent
function value as $f_i(t)$.  When making a prediction for a new
patient, we assume $S_i$ is known based on the patient's experience
before the procedure.  In the Supplementary Materials, we study the
robustness of our model to measurement error in $S_i$.

\subsection{Model Components}

As previously mentioned, we perform inference on recovery curves using
a hierarchical Bayesian model.  The Bayesian paradigm facilitates
sharing information across similar, but not identical patients.  This
information sharing is critical in our context as data on outcomes
after radical prostatectomy are very difficult to collect and
rare. The remainder of this subsection provides a detailed description
of the model.

Recall that our model is designed to be \emph{interpretable}.  In
particular, a patient's posterior over recovery curves firstly should only
have support over the space of recovery curves, so that a posterior as
in Figure \ref{fig:improved_curve_distribution} is not acceptable,
containing support over trajectories whose aymptotic level exceeds
that pre-treatment.  Secondly, the posterior should be unimodal, so
that a posterior as in Figure \ref{fig:bimodal_distribution} is not
acceptable.  We encourage the first desiderata by appropriately
constraining the support of relevant conditional distributions of the model,
and the second by guaranteeing unimodality of those conditional
distributions by using specialized distribution parameterizations.


First, recall the observed data $y_i(t)$ is a patient's \emph{reported} function level at time $t$.  We assume the reported function level comes from a likelihood that is a mixture distribution of the form:
\begin{align*}
  y_i(t)|f_i(t), \theta,p,\phi_M &\sim \theta \operatorname{bernoulli}(p) +
  (1-\theta)\operatorname{beta}_{m,\phi}(f_i(t), \phi_M),\ 
\  \text{with}\ 
p,\theta,\phi_M \in (0,1), X_i \in \mathbb{R}^K.
\end{align*}

We use a mixture distribution because patients can and do report values of 0 and 1.  In the data presented in Section~\ref{sec:real}, approximately five percent of patient responses are on the boundary of the unit interval.  The mixture distribution, therefore, places finite mass on the 0 and 1 responses, but also allows responses between 0 and 1 to be modeled using a recovery curve. 

For values other than 0 and 1, we propose a beta distribution that
depends on the patient's (latent) recovery curve value at time $t$,
$f_i(t)$.  Even notwithstanding the potential for values on the
boundary, parameterizing the unit interval in a way that is
interpretable is challenging.  To encourage unimodality of the $y_i(t)$, we would like
this beta distribution to always be unimodal.  For the typical
parameterization of the beta distribution,
$\operatorname{beta}_{\alpha, \beta}(\alpha',\beta')$), the mean
($\alpha'/\alpha'+\beta'$) and mode ($\alpha'-1/\alpha'+\beta'-1$)
both depend on both parameters.  Further, the distribution is only
unimodal if $\alpha$ and $\beta$ are both greater than one.  As our
goal is to develop a method that is easy to explain to clinicians
and patients, we chose to reparameterize the beta distribution in
terms of mode $m$ and spread parameter $\phi$.  This $\operatorname{beta}_{m,\phi}$ parameterization relates to the typical beta as:
\begin{equation}
\operatorname{beta}_{m,\phi}(m',\phi') =
\operatorname{beta}_{\alpha,\beta}\left(1 + \left(\frac{1}{\phi'} - 1\right)m',1 +
\left(\frac{1}{\phi'} - 1\right)\left(1-m'\right)\right).
\end{equation}  
Critically, our $\operatorname{beta}_{m,\phi}(m',\phi')$ distribution has mode $m'$
and for \emph{all} $m'$, is unimodal if and only if spread parameter $\phi' \in
(0,1)$.  Examples of such distributions and a full description of the steps to
reparameterize are in Figure
\ref{fig:beta_dists} and Section \ref{sec:reparam} of the
Supplementary Materials.

For values not on the boundary, each respondents' reported function
value comes from a beta distribution centered on their true (latent)
function value, $f_i(t)$, and with spread around that function value
determined by parameter $\phi_M$.  The patient's latent function value
takes the form of a recovery curve described in
Section~\ref{sec:recovery}.  Following the Bayesian paradigm, we specify prior distributions for the parameters of the recovery curve.  We expect that patients that are observably similar will have similar recovery trajectories, so we model the parameters of each patient's recovery curve as a function of observable covariates.  Recall that the recovery curve depends on individual specific parameters $A_i, B_i$ and $C_i$ controlling the asymptotic decrease in function post treatment, initial drop post treatment, and rate of recovery, respectively.  Since $A_i$ and $B_i$ are scaled to be consistent across patients, they have support on $(0,1)$.  The $C_i$ parameter is a rate and thus has support on $\mathbb{R}^+$.  We model each with a generalized linear model of the form:
\begin{align}
  A_i | b_A,\phi_A; z_A, X_i &\sim \operatorname{beta}_{m,\phi}(\operatorname{logistic}(z_A + b_A^T
  X_i), \phi_A)\label{eq:GLM_A}\\
  B_i |b_B, \phi_B; z_B, X_i &\sim \operatorname{beta}_{m,\phi}(\operatorname{logistic}(z_B + b_B^T
  X_i), \phi_B)\label{eq:GLM_B}\\
  C_i |b_C, \phi_C; z_C, X_i &\sim \operatorname{gamma}_{m,\phi}(\operatorname{exp}(z_C + b_C^T
  X_i), \phi_C),\ \text{with}\label{eq:GLM_C}\\
b_A,b_B,b_C & \in \mathbb{R}^K,\ 
\phi_A,\phi_B,\phi_C  \in (0,1), X_i \in \mathbb{R}^K.
\end{align}
Note that by assuming the recovery curve parameterization of Equation
\ref{eq:f}, we are effectively modeling a patient's function values \emph{scaled}
by their known pre-treatment value.  Justification  for this approach
is given in Section \ref{sec:scaled} of the Supplementary Materials.

To promote interpretability by encouraging unimodality of conditional
distributions, we again use the alternative parameterization of the
beta distribution described above for the likelihood.  Each patient's
initial drop in function value ($B_i$) is centered at a mode given by
the expected drop based on patients with similar observable
characteristics but with spread $\phi_B$.  A similar interpretation
applies to the eventual drop, $A_i$.  To ensure that $C_i$ is also
interpretable, we perform a similar reparameterization for the gamma
distribution.  A $\operatorname{gamma}_{m,\phi}(m',\phi')$ distribution has mode $m'$
and for \emph{all} $m'$, is unimodal if and only if spread parameter $\phi' \in
(0,1)$.  Examples of such distributions and details of the reparameterization
are in Figure
\ref{fig:gamma_dists} and Section \ref{sec:reparam} of the Supplementary Materials.  Under this
reparameterization the interpretation of the model for $C_i$ matches
$A_i$ and $B_i$.  The patient's rate of recovery is modeled using a
gamma distribution centered at the modal rate of observably similar
patients, with spread around that mode given by $\phi_C$.

The necessity for these specialized parameterizations of the beta and
gamma distributions becomes clear if
we consider a model that does not use them. Consider
the more traditional $\operatorname{beta}_{\mu,\beta}(\mu',\beta')$ parameterization,
where a $\operatorname{beta}_{\mu,\beta}(\mu',\beta')$ distribution has mean
$\mu'$, and $\beta'$ is a spread parameter.  Suppose we had let
$A_i|b_A,\phi_A;z_A,X_i \sim
\operatorname{beta}_{\mu,\beta}(\operatorname{logistic}(z_A + b_A
X_i), \phi_A)$.  A
$\operatorname{beta}_{\mu,\beta}(\mu',\beta')$ distribution is unimodal if and only
if $\beta'>1$ and $\mu'>\tfrac{1}{1+\beta'}$. 
Given $\beta'$, there is some $\mu'$ for which
a $\operatorname{beta}_{\mu,\beta}(\mu',\beta')$ distribution is not
unimodal.  Thus given $b_A$ and $\phi_A$, there would exist some
$X_i$ for which $A_i|b_A,\phi_A;z_A,X_i$ would not be
unimodal, which violates Property \ref{prop:unimodal}.  Similar reasoning applies to the gamma parameterization.

At this point only the prior distributions for the hyperparameters
must be specified to complete the model description.  
We encourage regularization on the regression coefficients by letting:\vspace{-0pt}

\begin{equation}
  \phi_A; \lambda_A \sim \operatorname{exp}(\lambda_A, 1),\ 
  \phi_B; \lambda_B \sim \operatorname{exp}(\lambda_B, 1),\ 
  \phi_C; \lambda_C \sim \operatorname{exp}(\lambda_C, 1)
\end{equation}where $\operatorname{exp}(\lambda,1)$ denotes an exponential
distribution with rate parameter $\lambda$ truncated on the right at
$1$ and $\lambda_A,\lambda_B,\lambda_C$ are hyperparameters.

Further, we assume there is some
``average'' recovery shape $g(\cdot;\mu_A,\mu_B,\mu_C)$ such that the
prior expected recovery curve of a
``average'' patient (one whose value of each feature is equal to the
mean of that feature in the dataset) is centered about
$S_ig(\cdot;\mu_A,\mu_B,\mu_C)$ (see Equation \ref{eq:shape}). 
That is, for the ``average'' patient, we want the conditional prior distributions of
$A_i,B_i,C_i$ to be centered at $\mu_A,\mu_B,\mu_C$,
respectively.  We will normalize all features to have mean 0 and
unit standard deviation, so that the ``average'' patient has a feature
vector consisting of all 0's.   Thus in light of Equations \ref{eq:GLM_A},
\ref{eq:GLM_B}, \ref{eq:GLM_C}, we let
\begin{align}
  \hspace*{-0pt}&z_A \sim \operatorname{normal}(\operatorname{logit}(\mu_A),s_A),\ 
  z_B \sim \operatorname{normal}(\operatorname{logit}(\mu_B),s_B),\ 
  z_C \sim \operatorname{normal}(\operatorname{exp}(\mu_C),s_C),\ \hspace*{-0pt} \text{and}\\
&b_A \sim \operatorname{multi\_normal}(\vec{0}, s_A I),\ 
b_B \sim \operatorname{multi\_normal}(\vec{0}, s_B I),\ 
b_C \sim \operatorname{multi\_normal}(\vec{0}, s_C I),\ 
\end{align}
\vspace{-0pt}where $\mu_A,\mu_B,\mu_C\in \mathbb{R},s_A,s_B,s_C \in \mathbb{R}^+$ are
hyperparameters, $\vec{0}$ is the $K$-dimensional 0 vector, and $I$ is
the $K$-dimensional identity matrix.  
Note that the
intercept for the regressions of Equations \ref{eq:GLM_A},
\ref{eq:GLM_B}, \ref{eq:GLM_C} is given a prior and not fixed.

Finally, without any prior belief about the parameters $p,\theta$
governing the likelihood, we let:
\begin{equation}
  p \sim \operatorname{unif}(0,1),\ 
  \theta \sim \operatorname{unif}(0,1).
\end{equation}

\subsection{Recap of Model features\label{subsec:requirements}}
We recap below the desired properties of our model, and how
our model satisfies those properties.

\vspace{-0pt}

\begin{enumerate}[leftmargin=*]
\itemsep -0em
  \item \label{prop:hierarchical} \textbf{Property:} Observed
    within-patient function values should be dependent, and post-treatment values for patients with similar
    covariates should be shrunk towards each other.  \\
    \textbf{Solution:} We adopted a hierarchical Bayesian model.  
Shrinkage was
   accomplished by letting $A_i$ be drawn from a single
   covariate dependent distribution.  In particular, $A_i$ was
   modelled using (a variant of) a generalized linear model.  An
   analogous approach models $B_i$ and $C_i$.
  \item \label{prop:realistic} \textbf{Property:} For the sake of
    interpretability, for each patient, their distribution over the underlying post-treatment
    function value should be a recovery curve -
    those functions satisfying requirements \ref{req:1} -
    \ref{req:4}.  A predictive distribution
    like that in Figure
    \ref{fig:improved_curve_distribution} is not acceptable.\\
    \textbf{Solution:} We respect the constraints of Equation \ref{req:7} in modelling
    $A_i, B_i, C_i$, letting their
    generalized linear models have beta, beta, and gamma
    response distributions, respectively, as these are canonical
    distributions with the desired support.
  \item \label{prop:unimodal}\textbf{Property:} For the sake of
    interpretability, we want the posterior of $f_i(t)$ to be unimodal.  For example,
    we do not want the predictive distribution to be bimodal, like that in
    Figure \ref{fig:bimodal_distribution}.\\
    \textbf{Solution:} 
The conditional distribution $A_i | b_A,
    \phi_A;X_i$ was constrained to be unimodal, for all $X_i$, $b_A$, and $\phi_A$.
    An analogous approach and constraint were used to model
    $B_i$ and $C_i$.  Ensuring this
    unimodality required special parameterizations of the beta and
    gamma distributions.
  \item \label{prop:likelihood} \textbf{Property:} $y_i(t)$ should have support on the
    \emph{closed} unit interval, because we observed that roughly 5\%
    of the time, patients recorded a 0 or 1 response.\\
    \textbf{Solution:} $y_i(t)$ comes from a mixture of a beta
    centered at $f_i(t)$ and a Bernoulli distribution.
  \item \label{prop:prior} \textbf{Property:} In the prior, a patient's distribution
    over recovery shapes should be
    centered about some ``average'' shape, given by curve parameters $\mu_A,\mu_B,\mu_C$.\\
    \textbf{Solution:} The GLM
    modeling $A_i$ depends on hyperparameter bias term $z_A$.
    $z_A$ was chosen so that in the prior, $A_i | B_A,
    \phi_A;X_i$ is centered at $\mu_A$.  Analogous approaches model $B_i$ and $C_i$.
\end{enumerate}

\section{Simulation Studies\label{sec:sim}}

Here, we examine
the ability of our model to recover the model parameters as the amount of
data simulated using those parameters grew.  
We chose a
single set of shared model parameters and hyperparameters
$\mu_A,\mu_B,\mu_C$.  Then, we performed the following for several
values of $N$, the number of patients in a simulated dataset:  We simulated
100 datasets, where for each dataset we used that set of chosen parameters to simulate observed function values
$y_i(t)$ for $N$ patients at times $t \in
\{1,2,4,8,12,18,24,30,36,42,48\}$,  the same times at which data were observed in the prostate cancer
dataset.  For each dataset, we obtained for each parameter a point
prediction as its posterior median, and calculated two quantities: the
signed error and unsigned error.  Figure \ref{fig:sim_study_vary_N_signed} shows the mean and
standard deviation of the signed error across the 100 datasets for each
parameter, for various values of N (N$\in\{50,100,250,500,1000,2500,5000\}$).  Please see Figure \ref{fig:sim_study_vary_N_unsigned} of the
Supplementary Materials for the analogous
information, for the unsigned error.  Note that Figure
\ref{fig:sim_study_vary_N_signed} thus shows the bias and variance
of the point estimates of the parameters, with the estimator being their
respective posterior medians.


The set of parameters we used was simply one that was not pathological.  We used
$b_A=1, b_B=2, b_C=3, \theta=0.1, p=0.3,
\phi_A=\phi_B=\phi_C=\phi_M=0.01$, and
$\mu_A=0.4,\mu_B=0.7,\mu_C=5$.  We assume only 1 feature, which for each sample is
generated from a unit normal distribution. For inference, we set $s_A
= s_B = s_C = 1$, $\lambda_A = \lambda_B = \lambda_C =10$ and
$\lambda_M =10$.  To obtain posterior samples, we used Stan
\citep{Hoffman2011}, obtaining 2500 samples from each of 4 chains with
no thinning, using 2500 burn-in steps.  We assessed
convergence both by using the Gelman statistic
\citep{AndrewGelman1992} and visual examination of the traces for each
parameter.
We checked that in fitting the model to each simulated
dataset, the maximum Gelman statistic over parameters was less than
$1.2$.  The meaning of errors in the regression parameter is
provided by Equations \ref{eq:GLM_A}-\ref{eq:GLM_C}, and the meaning
of errors in the
spread parameters $\phi_A,\phi_B,\phi_C,\phi_M$ is provided by the
plots of beta and gamma distributions in Figures \ref{fig:beta_dists}
and \ref{fig:gamma_dists} of the Supplementary Materials.



\section{Analysis of Prostate Cancer Dataset\label{sec:real}}
\subsection{Dataset Description\label{sec:data}}
Our data comes from a study \citep{Gore2009,Gore2010} that
prospectively tracked the sexual function as measured using the UCLA
Prostate Cancer Index \citep{Litwin1998} of 304 patients who underwent
radical prostatectomy to treat clinically localized prostate
cancer. After applying dataset filters as detailed in Section \ref{sec:filtering} of the
Supplementary Materials, data from 237 patients is retained.  Their sexual function levels were collected right before
treatment and over a 48-month post-treatment study period via mailed surveys at
1,2,4,8,12,18,24,30,36,42, and 48 months after their respective
treatments, and missing data was due to lack of survey response.
The Prostate Cancer Index, derived from answers to a series of
multiple choice questions, is a numerical measure of a patient's level
of sexual function that lies between 0 and 100, which we scale to the
unit interval.  Various patient covariates were collected at time of treatment, including age, cancer grade/stage, physical/mental condition, uninary/bowel function, and comorbidity count.

Prostate cancer patients' post-prostatectomy sexual function outcomes can be modulated
by non-mandatory post-prostectomy treatments such as the use of an
erectile aid.  As such additional treatments are non-standard, our
goal in this particular analysis is to model the sexual function
outcomes for patients who would not receive them.  Furthermore, we are
not interested in modeling the post-prostatectomy sexual function of
patients whose sexual function prior to the potential prostatectomy is already close to 0, as
such patients' post-treatment sexual function would be expected to
remain constant afterwards, following a different model that is uninteresting to
analyze.  
Thus, we define the
target population of this model to be patients 
considering a prostatectomy, who satisfy the following two
properties:  firstly, they would not use any additional remedial treatments
post-prostatectomy, such as an erectile aid, and secondly, they would have a
non-neglible level of sexual function prior to receiving the potential
prostatectomy.  The dataset filters we applied retain members of this
target population.

\subsection{Choosing Features\label{sec:covariates}}

To identify potential correlates of recovery curve shapes, for every
patient, we used curve fitting
to find the $A,B,C$ parameters corresponding to their post-event
recovery shapes.  We made scatter plots of each of those parameters
against all available covariates to identify ones that correlated with
curve parameters, and identified the pre-treatment sexual function
level (referred to as ``init'' in all figures) and patient age (at
treatment time) to be the 2 covariates most strongly correlated with
curve parameters.  From the scatter plots (in
Figure~\ref{fig:scatters} of the Supplementary Materials), we saw the relationship between
those 2 covariates and curve parameters is likely nonlinear.  Thus we
created binned categorical features based on age and pre-treatment
function level.  The bins for the categorical features we used were as follows:
\begin{itemize}
\item age (in years): 0 to 55, 55 to 65, 65+
\item pre-treatment function level: 0 to 41, 41 to 60, 60 to 80, 80 to 100
\end{itemize}
We note that such subdivisions matches with the urologist co-author's
clinical experience regarding how urologists categorize age and
pre-treatment function level.  

Thus, in our model, patients belong to 1 of 12 classes, depending on
into which of the three age groups they fall into, and which of the
four intervals their pre-treatment sexual function level lies.  These
features were normalized.  To visualize the effect of these 2
covariates on recovery shape from another view, we stratified the
patients by age category and pre-treatment sexual function level
category, and plotted (see Figure \ref{fig:stratified_real_data} of
the Supplementary Materials) the average shape of the patients in each category.

\vspace*{-0pt}
\subsection{Fitting Our Model\label{sec:apply}}
Now, we describe how we chose hyperparameters, and the fitting of the model.
To choose $\mu_A, \mu_B, \mu_C$, which describe the average recovery
shape for the ``average'' patient in the target population, we fit a recovery shape using our
parametric form to the training fold-wide average scaled function
value shown in Figure \ref{fig:avg_curve} (labeled ``average
shape'').  The values we used for the remaining hyperparameters were
$s_A=s_B=s_C=1$ and $l_A=l_B=l_C=l_M=10$.  We show in Section \ref{sec:hyper_performance} of the Supplementary Materials that
out-of-sample performance (as described in Section \ref{sec:performance}) is not sensitive to the particular choice of
those hyperparameters.  

To fit the
model, we used Stan \citep{Hoffman2011}, for each of 4 chains, running
2500 steps with 2500 burn-in steps and no thinning, and assessed
convergence using the Gelman statistic \citep{AndrewGelman1992} (The
maximum value of the Gelman statistic over all parameters was 1.11).  Please see
Section \ref{sec:checks} of the Supplementary Materials for posterior predictive checks.


\subsection{Out-of-Sample Performance\label{sec:performance}}
We measure the performance of our model by its ability to predict
$y_i(t)$, the observed function values.  We obtain a point prediction
of $y_i(t)$, denoted $\hat{y}_i(t)$, via the median of the posterior
distribution of $f_i(t)$, the ``underlying'' function value.  The loss function we use to measure
performance was absolute prediction error: the absolute
difference between $y_i(t)$ and $\hat{y}_i(t)$.   
To measure out-of-sample performance, we performed 5-fold
cross-validation, obtaining, for each test sample, point predictions
from our model, and examined the average, over
the test folds, of loss at a given time as measured by absolute prediction error. (The
average loss at time $t$ for a test fold consisting of the patient index
set $I$ for which function values were recorded at time $t$ is
$\tfrac{1}{|I|} \sum_{i \in I}|\hat{y}_i(t) - y_i(t)|$ where
$\hat{y}_i(t)$ is the point prediction of the function level of
patient $i$ at time $t$ and $|I|$ is the size of index set $I$.)  In
particular, the entire time series for patients in the testing folds
are predicted given the entire time series of the patients in the training fold.  Data from the early part of one patient's€™ time series is
\emph{not} used to predict the same patient'€™s future values.  We plot, over
time, the out-of-sample performance of our model, as well as that of
two baseline models in Figure \ref{fig:performance}.  Note that all comparison models, like our model, first predict the patient's \emph{scaled} function values, and then multiply it by their pre-treatment value to obtain a prediction of \emph{absolute} function value.  To compare
the improvement of our method to the status quo, in which a doctor
merely tells a patient the population-wide average shape, we plotted
the performance of simply predicting a patient to have the average
recovery shape, labelled ``mean''.  We compared the performance of our model
to a timewise scaled regression that, at each of the 11 common timepoints, uses a separate
generalized linear regression model to relate the scaled function
value at the timepoint to patient features. This model, labelled
``scaled regression'', uses a
logistic inverse link function and assumes a normal response
distribution. Finally, because of the high variance in our data, we
show for comparison the \emph{in-sample} performance of a model that is prone to overfitting. This model,
labelled ``median'', in order to make a
prediction for a patient at a given time, looks at which of the 12 patient classes the patient belongs
to and then calculates the median scaled function value, at the given time, of
patients who belong to that patient class, over the \emph{entire}
dataset.  
As can be seen in Figure \ref{fig:performance}, the out-of-sample
performance of our model is roughly equivalent to that of the scaled
regression model, though our model is more interpretable.  Error bars show the variance in estimates of the expected loss at each time.

\subsection{Interpretability of Model\label{sec:interp}}

Our model achieves out-of-sample performance comparable to that of the
timewise scaled regression described in Section \ref{sec:performance}.
However, our model produces much more easily understood predictions,
outputting a distribution over time series consisting solely of
recovery curves, so that they are smoothly increasing monotonically
towards an asymptote, and do not exceed the pre-treatment value.  In contrast, the timewise scaled regression model produces a time
series that is not guaranteed to be smooth or monotonically
increasing.  In addition to matching prior expectations, our model's
predictions are more quickly processed by the patient, which, as our
references to interpretability indicated, are crucial in a clinical
setting.  To
illustrate, in Figure \ref{fig:curve_comparisons} for several of the
12 classes of patients, we plot the scaled function values produced by
scaled timewise regression, the distribution over $f_t$ from our
model, and the timewise median of that distribution.  A patient,
expecting to see a recovery curve, can with a quick glance of the red
curves (timewise median of our predictions), pick up
what the initial drop, asymptotic drop, and recovery rate of their
predicted recovery curve are.  On the other hand, with the scaled timewise regression, the patient tries to
extrapolate what those same quantities are from the jagged predictions, finds it hard to do so, wondering whether the fluctuations are a real trend or
just noise.

Furthermore, we have designed our model so that prediction \emph{uncertainty} is easily interpretable when a patient's posterior distribution of curves is plotted.  Because we encourage the a patient's recovery curve parameter distribution to be unimodal in the posterior, we expect the pointwise distribution of curve values, namely that of $f_t$, to be unimodal.  This is why in Figure \ref{fig:curve_comparisons}, the distribution of curves appears clustered about the red curve.  It is important that one can visually extract from a plot of posterior distribution of curves a \emph{single} most likely curve.  Then, such a plot can be interpreted as giving a single curve prediction, along with the uncertainty in that prediction.  On the other hand, if the posterior distribution of curves were clustered around, say, two curves, there would be no such clear interpretation.  

\subsection{Dependence of Recovery Shape on Covariates and Comparison with Literature}

Our analysis teases apart the dependence of recovery shape on age and
pre-treatment value. 
In Figure \ref{fig:age_trend}, we examine the effect of patient age on
recovery curve shape by stratifying those curves by pre-treatment level.  We find when pre-treatment level is
controlled for, patients younger than 55 years of age have a smaller
asymptotic drop in sexual function level, proportional to their
pre-treatment level.  (We performed a one-sided z-test that the scaled
function value at 48 months for patients younger than 55 years of age was larger
than those not; p-value =
$.005$.) This effect is diminished for patients with pre-treatment level higher than 0.80.  When pre-treatment level is not controlled for, the asymptotic proportional drop in function level for younger patients is lower.  In both cases, the proportional initial drop in function level does not depend on age.
In Figure \ref{fig:init_trend} we examine the effect of pre-treatment sexual function level on recovery shape by stratifying those curves by age.  We find that when age is controlled for, patients with pre-treatment level higher than 0.80 have a smaller asymptotic drop in function level, proportional to their pre-treatment level. (One sided z-test p-value
= \num{4.17e-7}.) However, this effect is diminished for patients younger than 55.  The proportional initial drop in function depends mildly on pre-treatment level.  



Unlike past methods, which have mostly focused on modeling a
continuous or binary measure of sexual function at a single fixed time, our model makes predictions of the entire post-treatment function trajectory.  Regardless, we can still compare our findings to them.  Past work that modeled a continuous measure of sexual function found that lower age and higher pre-treatment sexual function level are statistically linked to higher \emph{absolute} levels of that measure \citep{Talcott2003}, and that lower age is linked to a smaller \emph{change} in that measure of function level \citep{Vordermark2008}.  Likewise, when a binary indicator of satisfactory sexual function has been logistically regressed against patient covariates, lower age \citep{Regan2011, Ayyathurai2008} and higher pre-treatment function level (\citeauthor{Regan2011}) have been found to lead to a higher probability of having satisfactory sexual function.  One can conclude from these past statistical analyses, as well as model-free data analyses \citep{Rabbani2000, Michl2006}, that lower age and higher pre-treatment sexual function level, by any measure, are linked to higher post-treatment sexual function level, agreeing with our findings.
Though, we stress that unlike any previous analysis, we model the link between patient features and longitudinal sexual function levels \emph{proportional} to the pre-treatment level.  

\vspace*{-0pt}

\section{Conclusion\label{sec:conclusion}}
We presented a Bayesian model that can be used to predict recovery curves, which arise in many medical contexts. Our overarching goal is to facilitate the flow of information from the data to the user, who may not be statistically inclined. Towards this end, we impart interpretability to both the model and its output, a model that is easily explained and produces believable outputs is more clinically applicable.
In particular, our model predicts quantities that are of natural interest, and guarantees that its output is in fact the recovery curve that we assume a domain expert to expect of a prediction. Furthermore, our model is designed for easy visualization of predictions and the associated uncertainty, as we encourage the posterior distribution over recovery curves to have a clear 'mode'.
We used our model to analyze the impact of prostatectomy on a patient's post-treatment sexual function trajectory, and characterized the extent of that impact on patient age and pre-treatment sexual function level, producing conclusions that agree with and supplement past findings.  We believe our model can provide insights in other medical domains as well.

\section{Supplementary Materials}
\label{sec6}

The Supplementary Materials are available at
\url{http://biostatistics.oxfordjournals.org}.  It contains details of
the reparameterizations of the beta and gamma distributions,
exploratory plots and details of filtering of the prostatectomy
dataset, as well as an additional simulation study.  Furthermore, it
contains further analysis of the prostatectomy dataset as it relates
to our model: posterior predictive checks and sensitivity of results to hyperparameters
and measurement error in the pre-treatment function level.  Finally,
it contains high resolution plots of the posterior predictive
distribution for the patients in the prostatectomy dataset.  Simulated data and code are
available at \url{https://github.com/fultonwang/recovery\_curve}.
\section{Acknowledgements}
This work was supported by National Science Foundation CAREER grant
IIS-1053407 to C. Rudin and National Science Foundation grant DMS-1737673 to T. McCormick.  We thank Dr. Jim Michaelson of Massachusetts General Hospital for helpful discussions.

\bibliographystyle{biorefs}
\bibliography{prostate}

\begin{thebibliography}{99}

\bibitem[Ayyathurai \emph{and others}(2008)Ayyathurai, Manoharan, Nieder, Kava
  and Soloway]{Ayyathurai2008}
\textsc{Ayyathurai, R., Manoharan, M., Nieder, A.~M., Kava, B. and Soloway,
  M.~S.} (2008).
\newblock {Factors affecting erectile function after radical retropubic
  prostatectomy: results from 1620 consecutive patients.}
\newblock {\em BJU international\/}~\textbf{101}(7), 833--6.

\bibitem[Briganti \emph{and others}(2011)Briganti, Gallina, Suardi, Capitanio,
  Tutolo, Bianchi, Salonia, Colombo, {Di Girolamo}, Martinez-Salamanca,
  Guazzoni, Rigatti and Montorsi]{Briganti2011}
\textsc{Briganti, A., Gallina, A., Suardi, N., Capitanio, U., Tutolo, M.,
  Bianchi, M., Salonia, A., Colombo, R., {Di Girolamo}, V., Martinez-Salamanca,
  J.~I., Guazzoni, G., Rigatti, P.} \emph{and others}. (2011).
\newblock {What is the definition of a satisfactory erectile function after
  bilateral nerve sparing radical prostatectomy?}
\newblock {\em The Journal of Sexual Medicine\/}~\textbf{8}(4), 1210--7.

\bibitem[Cai and Dunson(2007)Cai and Dunson]{Cai2007}
\textsc{Cai, B. and Dunson, D.~B.} (2007).
\newblock {Bayesian multivariate isotonic regression splines}.
\newblock {\em Journal of the American Statistical
  Association\/}~\textbf{102}(480), 1158--1171.

\bibitem[Denison \emph{and others}(1998)Denison, Mallick and
  Smith]{Denison1998}
\textsc{Denison, D. G.~T., Mallick, B.~K. and Smith, A. F.~M.} (1998).
\newblock {Automatic Bayesian curve fitting}.
\newblock {\em Journal of Royal Statistical Society Series B\/}.

\bibitem[Descazeaud \emph{and others}(2006)Descazeaud, Debr\'{e} and
  Flam]{Descazeaud2006}
\textsc{Descazeaud, A., Debr\'{e}, B. and Flam, T.~A.} (2006).
\newblock {Age difference between patient and partner is a predictive factor of
  potency rate following radical prostatectomy}.
\newblock {\em The Journal of Urology\/}~\textbf{176}(6), 2594--2598.

\bibitem[Eastham \emph{and others}(2008)Eastham, Scardino and
  Kattan]{Eastham2008}
\textsc{Eastham, J.~A., Scardino, P.~T. and Kattan, M.~W.} (2008).
\newblock {Predicting an optimal outcome after radical prostatectomy: the
  trifecta nomogram}.
\newblock {\em The Journal of Urology\/}~\textbf{179}(6), 2207--2211.

\bibitem[Gelman and Rubin(1992)Gelman and Rubin]{AndrewGelman1992}
\textsc{Gelman, A. and Rubin, D.} (1992).
\newblock {Inference from iterative simulation using multiple sequences}.
\newblock {\em Statistical Science\/}~\textbf{7}(4), 457--511.

\bibitem[Gore \emph{and others}(2010)Gore, Gollapudi, Bergman, Kwan, Krupski
  and Litwin]{Gore2010}
\textsc{Gore, J.~L., Gollapudi, K., Bergman, J., Kwan, L., Krupski, T.~L. and
  Litwin, M.~S.} (2010).
\newblock {Correlates of bother following treatment for clinically localized
  prostate cancer.}
\newblock {\em The Journal of Urology\/}~\textbf{184}(4), 1309--15.

\bibitem[Gore \emph{and others}(2009)Gore, Kwan, Lee, Reiter and
  Litwin]{Gore2009}
\textsc{Gore, J.~L., Kwan, L., Lee, S.~P., Reiter, R.~E. and Litwin, M.~S.}
  (2009).
\newblock {Survivorship beyond convalescence: 48-month quality-of-life outcomes
  after treatment for localized prostate cancer.}
\newblock {\em Journal of the National Cancer Institute\/}~\textbf{101}(12),
  888--92.

\bibitem[Hartzler \emph{and others}(2015)Hartzler, Izard, Dalkin, Mikles and
  Gore]{hartzler2015design}
\textsc{Hartzler, A.~L., Izard, J.~P., Dalkin, B.~L., Mikles, S.~P. and Gore,
  J.~L.} (2015).
\newblock Design and feasibility of integrating personalized pro dashboards
  into prostate cancer care.
\newblock {\em Journal of the American Medical Informatics Association\/}.

\bibitem[Hoffman and Gelman(2011)Hoffman and Gelman]{Hoffman2011}
\textsc{Hoffman, M.~D. and Gelman, A.} (2011).
\newblock {The No-U-Turn sampler: adaptively setting path lengths in
  Hamiltonian Monte Carlo}.
\newblock {\em Journal of Machine Learning Research\/}, 30.

\bibitem[Jung and Wickrama(2008)Jung and Wickrama]{Jung2008}
\textsc{Jung, T. and Wickrama, K.} (2008).
\newblock {An introduction to latent class growth analysis and growth mixture
  modeling}.
\newblock {\em Social and Personality Psychology Compass\/}~\textbf{2}(1),
  302--317.

\bibitem[Litwin \emph{and others}(1998)Litwin, Hays, Fink, Ganz, Leake and
  Brook]{Litwin1998}
\textsc{Litwin, M.~S., Hays, R.~D., Fink, A., Ganz, P.~A., Leake, B. and Brook,
  R.~H.} (1998).
\newblock {The UCLA Prostate Cancer Index: development, reliability, and
  validity of a health-related quality of life measure.}
\newblock {\em Medical Care\/}~\textbf{36}(7), 1002--12.

\bibitem[Mammen(1991)Mammen]{Mammen}
\textsc{Mammen, E.} (1991).
\newblock {Estimating a smooth monotone regression function}.
\newblock {\em The Annals of Statistics\/}~\textbf{19}(2), 724--740.

\bibitem[Michl \emph{and others}(2006)Michl, Friedrich, Graefen, Haese, Heinzer
  and Huland]{Michl2006}
\textsc{Michl, U. H.~G., Friedrich, M.~G., Graefen, M., Haese, A., Heinzer, H.
  and Huland, H.} (2006).
\newblock {Prediction of postoperative sexual function after nerve sparing
  radical retropubic prostatectomy.}
\newblock {\em The Journal of Urology\/}~\textbf{176}(1), 227--31.

\bibitem[Nayak \emph{and others}(2015)Nayak, Hartzler, Macleod, Izard, Dalkin
  and Gore]{nayak2015relevance}
\textsc{Nayak, J.~G., Hartzler, A.~L., Macleod, L.~C., Izard, J.~P., Dalkin,
  B.~M. and Gore, J.~L.} (2015).
\newblock Relevance of graph literacy in the development of patient-centered
  communication tools.
\newblock {\em Patient Education and Counseling\/}.

\bibitem[Neelon and Dunson(2004)Neelon and Dunson]{Neelon2004}
\textsc{Neelon, B. and Dunson, D.~B.} (2004).
\newblock {Bayesian isotonic regression and trend analysis.}
\newblock {\em Biometrics\/}~\textbf{60}(2), 398--406.

\bibitem[Potosky \emph{and others}(2004)Potosky, Davis, Hoffman, Stanford,
  Stephenson, Penson and Harlan]{Potosky2004}
\textsc{Potosky, A.~L., Davis, W.~W., Hoffman, R.~M., Stanford, J.~L.,
  Stephenson, R., Penson, D.~F. and Harlan, L.~C.} (2004).
\newblock {Five-year outcomes after prostatectomy or radiotherapy for prostate
  cancer: The Prostate Cancer Outcomes Study.}
\newblock {\em Journal of the National Cancer Institute\/}~\textbf{96(18)}(18),
  1358--67.

\bibitem[Rabbani \emph{and others}(2000)Rabbani, Stapleton, Kattan, Wheeler and
  Scardino]{Rabbani2000}
\textsc{Rabbani, F., Stapleton, A.~M., Kattan, M.~W., Wheeler, T.~M. and
  Scardino, P.~T.} (2000).
\newblock {Factors predicting recovery of erections after radical
  prostatectomy.}
\newblock {\em The Journal of Urology\/}~\textbf{164}(6), 1929--34.

\bibitem[Ramsay and Silverman(2002)Ramsay and Silverman]{Ramsay2002}
\textsc{Ramsay, J.~O. and Silverman, B.~W.} (editors).  (2002).
\newblock {\em {Applied functional data analysis: methods and case studies}\/},
  Springer Series in Statistics. New York, NY.

\bibitem[Regan \emph{and others}(2011)Regan, Cooperberg, Wei, Michalski,
  Sandler, Litwin, Klein, Kibel, Hamstra, Pisters, Kuban, Kaplan, Wood, Ciezki,
  Dunn, Carroll and Sanda]{Regan2011}
\textsc{Regan, M.~M., Cooperberg, M.~R., Wei, J.~T., Michalski, J.~M., Sandler,
  H.~M., Litwin, M.~S., Klein, E., Kibel, A.~S., Hamstra, D.~A., Pisters,
  L.~L., Kuban, D.~A., Kaplan, I.~D., Wood, D.~P., Ciezki, J., Dunn, R.~L.,
  Carroll, P.~R.} \emph{and others}. (2011).
\newblock {Prediction of erectile function following treatment for prostate
  cancer}.
\newblock {\em Journal of the American Medical Association\/}~\textbf{306}(11),
  1205--1214.

\bibitem[Rogosa and Willett(1985)Rogosa and Willett]{Rogosa1985}
\textsc{Rogosa, D.~R. and Willett, J.~B.} (1985).
\newblock {Understanding correlates of change by modeling individual
  differences in growth}.
\newblock {\em Psychometrika\/}~\textbf{50}(2), 203--228.

\bibitem[Rolfe \emph{and others}(2011)Rolfe, Mengersen, Vearncombe, Andrew and
  Beadle]{Rolfe2011}
\textsc{Rolfe, M.~I., Mengersen, K.~L., Vearncombe, K.~J., Andrew, B. and
  Beadle, G.~F.} (2011).
\newblock {Bayesian estimation of extent of recovery for aspects of verbal
  memory in women undergoing adjuvant chemotherapy treatment for breast
  cancer}.
\newblock {\em Journal of the Royal Statistical Society: Series C (Applied
  Statistics)\/}~\textbf{60}(5), 655--674.

\bibitem[Sanda \emph{and others}(2008)Sanda, Dunn, Michalski, Sandler,
  Northouse, Hembroff, Lin, Greenfield, Litwin, Saigal, Mahadevan, Klein,
  Kibel, Pisters, Kuban, Kaplan, Wood, Ciezki, Shah and Wei]{Vordermark2008}
\textsc{Sanda, M.~G, Dunn, R.~L., Michalski, J., Sandler, H.~M., Northouse, L.,
  Hembroff, L., Lin, X., Greenfield, T.~K., Litwin, M.~S., Saigal, C.~S.,
  Mahadevan, A., Klein, E., Kibel, A., Pisters, L.~L., Kuban, D., Kaplan, I.,
  Wood, D., Ciezki, J., Shah, N.} \emph{and others}. (2008).
\newblock {Quality of life and satisfaction with outcome among prostate-cancer
  survivors}.
\newblock {\em The New England Journal of Medicine\/}~\textbf{358}(12),
  1250--1261.

\bibitem[Shively \emph{and others}(2009)Shively, Sager and Walker]{Shively2009}
\textsc{Shively, T.~S., Sager, T.~W. and Walker, S.~G.} (2009).
\newblock {A Bayesian approach to non-parametric monotone function estimation}.
\newblock {\em Journal of the Royal Statistical Society: Series B (Statistical
  Methodology)\/}~\textbf{71}(1), 159--175.

\bibitem[Talcott \emph{and others}(2003)Talcott, Manola, Clark, Kaplan, Beard,
  Mitchell, Chen, O'Leary, Kantoff and D'Amico]{Talcott2003}
\textsc{Talcott, J.~A., Manola, J., Clark, J.~A., Kaplan, I., Beard, C.~J.,
  Mitchell, S.~P., Chen, R.~C., O'Leary, M.~P., Kantoff, P.~W. and D'Amico,
  A.~V.} (2003).
\newblock {Time course and predictors of symptoms after primary prostate cancer
  therapy}.
\newblock {\em Journal of Clinical Oncology\/}~\textbf{21}(21), 3979--86.

\bibitem[Tilling \emph{and others}(2001)Tilling, Sterne and
  Wolfe]{Tilling2001a}
\textsc{Tilling, K., Sterne, J. A.~C. and Wolfe, C. D.~A.} (2001).
\newblock {Multilevel growth curve models with covariate effects : application
  to recovery after stroke}.
\newblock {\em Statistics in Medicine\/}~\textbf{20}(5), 685--704.

\bibitem[Warschausky \emph{and others}(2001)Warschausky, Kay and
  Kewman]{Warschausky2001}
\textsc{Warschausky, S., Kay, J.~B. and Kewman, D.~G.} (2001).
\newblock {Hierarchical linear modeling of FIM instrument growth curve
  characteristics after spinal cord injury.}
\newblock {\em Archives of Physical Medicine and
  Rehabilitation\/}~\textbf{82}(3), 329--34.

\end{thebibliography}




\newpage

\begin{figure}
\centering
\begin{subfigure}{0.48\linewidth}
\centering 
\includegraphics[width=1\linewidth]{\getFullPath{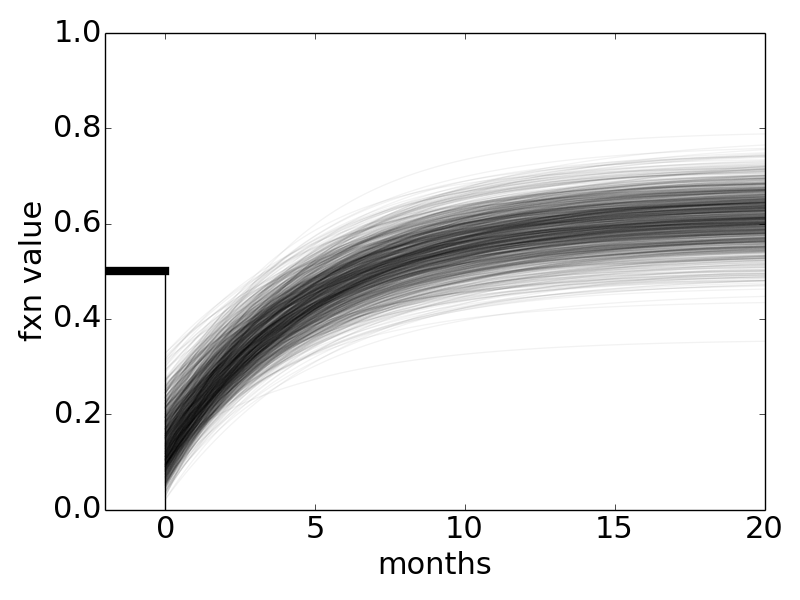}}
\caption{This predictive distribution is unrealistic because some of
  the time series are not recovery curves, as their post-treatment
  function value exceeds that pre-treatment.\\ \label{fig:improved_curve_distribution}}
\label{fig:improved_curve_distribution}
\end{subfigure}\hspace{1em}
\begin{subfigure}{0.48\linewidth}
\centering \includegraphics[width=1\linewidth]{\getFullPath{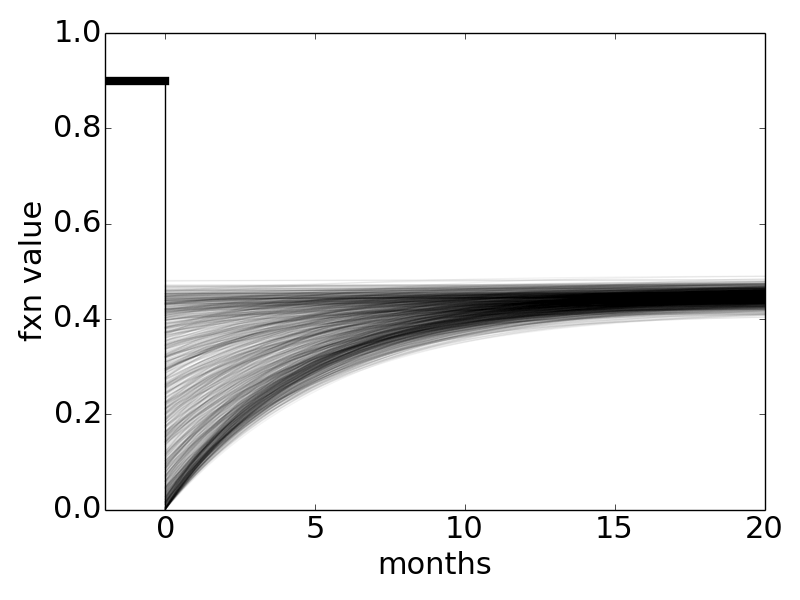}}
\caption{This predictive distribution is unrealistic
  because the distribution is not unimodal. One can see this because there are two dark sets of curves, one for each mode of the posterior curve distribution. \label{fig:bimodal_distribution}}
\label{fig:bimodal_distribution}
\end{subfigure}
\caption{Two unrealistic predictive distributions}
\label{fig:unrealistic}
\end{figure}

\begin{figure}
\centering
\begin{subfigure}{0.5\linewidth}
\centering 
\includegraphics[width=1\linewidth]{\getFullPath{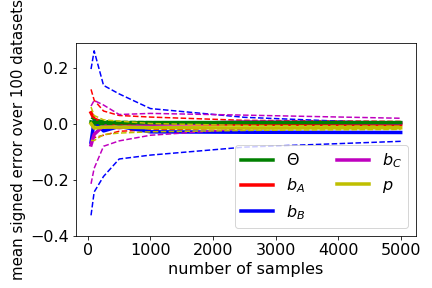}}
\end{subfigure}
\begin{subfigure}{0.5\linewidth}
\centering \includegraphics[width=1\linewidth]{\getFullPath{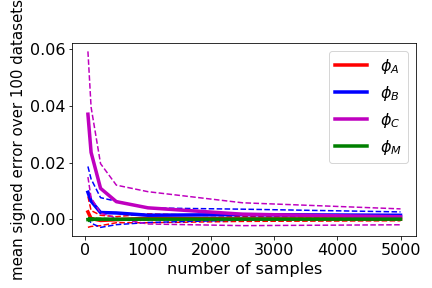}}
\end{subfigure}
\caption{For each parameter, the mean signed error over the
  simulated datasets decreases with the size of the simulated
  datasets.  Dotted lines denote 1 standard deviation.}
\label{fig:sim_study_vary_N_signed}
\end{figure}


\begin{figure}
\centering
\begin{subfigure}{0.48\linewidth}
\centering 
\includegraphics[width=1\linewidth]{\getFullPath{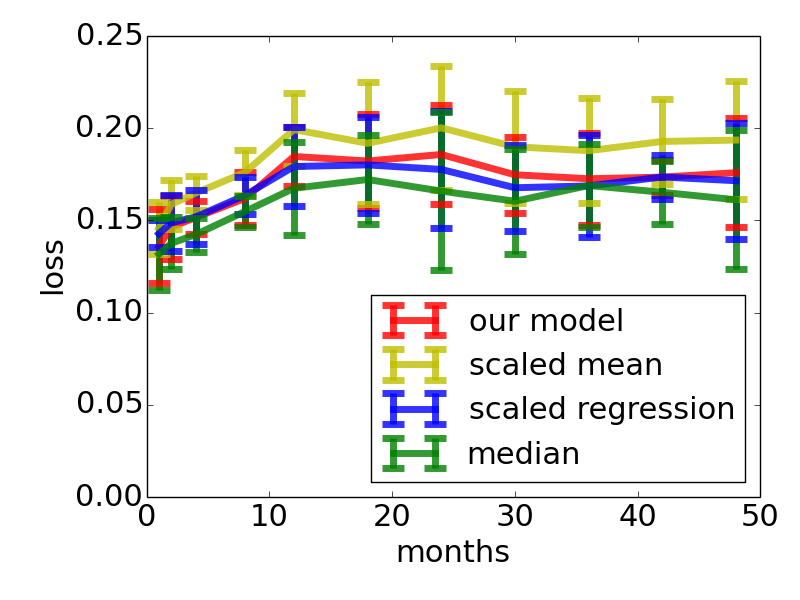}}
\caption{The out-of-sample performance of our model is comparable to
  that of timewise scaled regression, and approaches the
  \emph{in-sample} performance of
  ``median'', a method prone to overfitting.}
\label{fig:performance}
\end{subfigure}\hspace{1em}
\begin{subfigure}{0.48\linewidth}
\centering
\includegraphics[width=1\linewidth]{\getFullPath{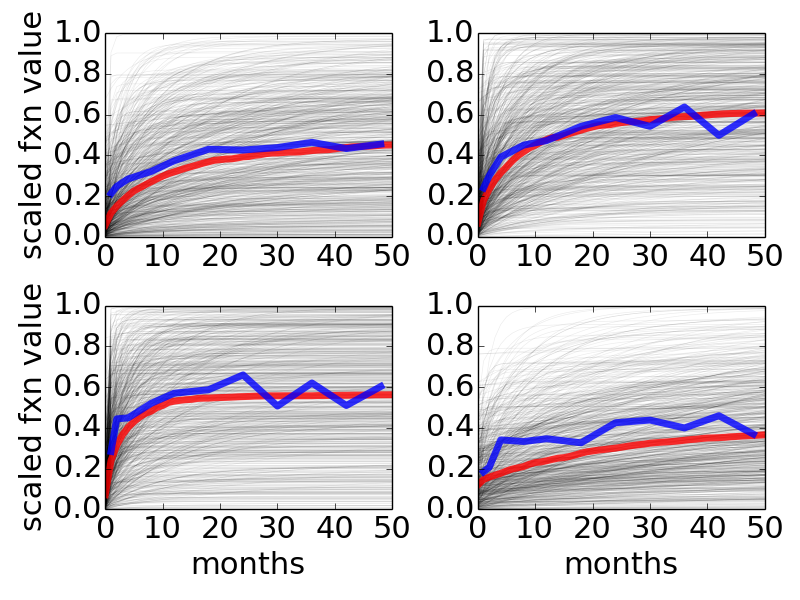}}
\caption{The posterior predictive distribution over recovery curves (black) and timewise medians of it (red) convey more plausible predictions than that of timewise scaled regression (blue), whose prediction is not guaranteed to be a recovery curve.}
\label{fig:curve_comparisons}
\end{subfigure}
\end{figure}

\begin{figure}
\centering
\begin{subfigure}{0.5\linewidth}
\centering 
\includegraphics[width=1\linewidth]{\getFullPath{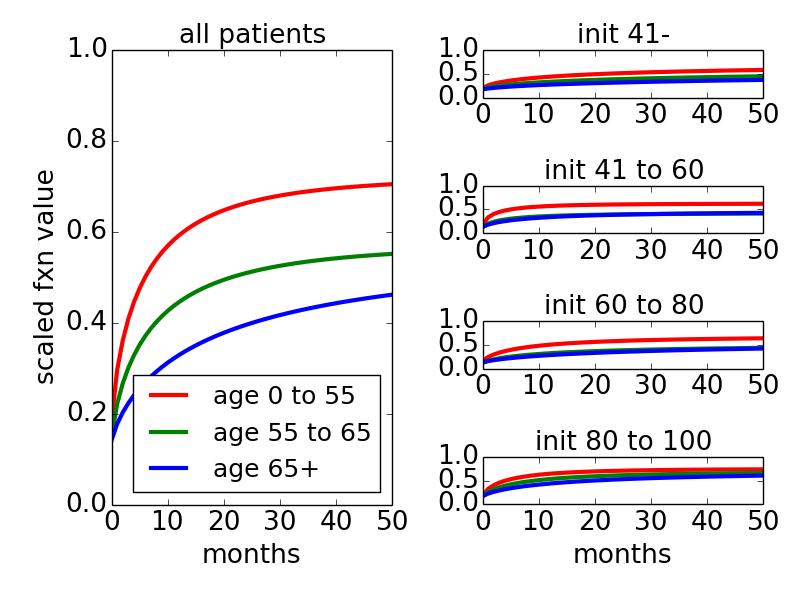}}
\caption{age trend}
\label{fig:age_trend}
\end{subfigure}
\begin{subfigure}{0.5\linewidth}
\centering \includegraphics[width=1\linewidth]{\getFullPath{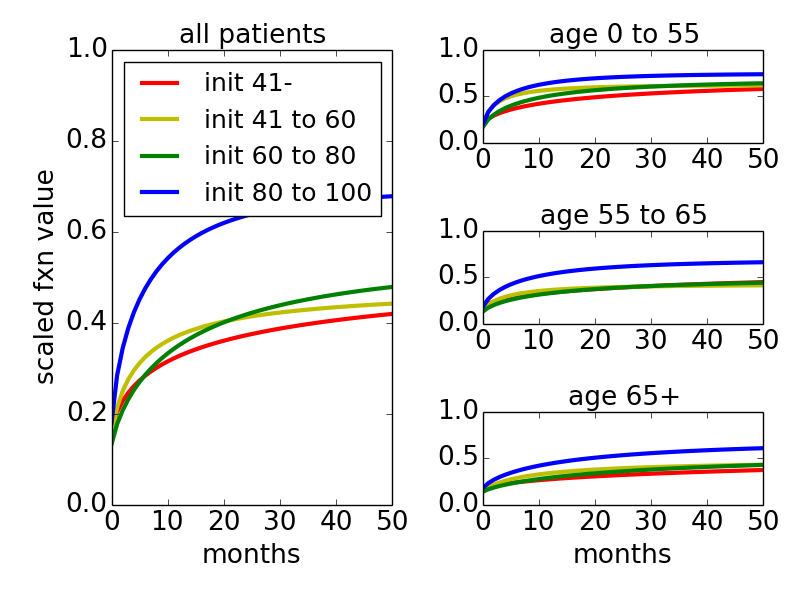}}
\caption{pre-treatment level trend}
\label{fig:init_trend}
\end{subfigure}
\caption{Our model identifies the relationship between pre-treatment level, age, and a patient's latent recovery shape.}
\label{fig:trends}
\end{figure}


\clearpage
\title{Supplementary Material}
\maketitle
\setcounter{section}{0}
\section{Reparameterization of beta and gamma distributions\label{sec:reparam}}
Our $\operatorname{beta}_{m,\phi}(\cdot,\cdot)$ parameterization is
based on the traditional
$\operatorname{beta}_{\alpha,\beta}(\cdot,\cdot)$ parameterization.  A
$\operatorname{beta}_{\alpha,\beta}(\alpha',\beta')$ distribution is
unimodal with mode $\frac{\alpha'-1}{\alpha'+\beta'-1}$ if and only if
$\alpha'>1, \beta'>1$.  Making the substitution $\alpha = 1 + sm$ and
$\beta = 1 + s(1-m)$, we obtain the parameterization
$\operatorname{beta}_{m,s}(m',s') =
\operatorname{beta}_{\alpha,\beta}(1 + s'm',1 + s'(1-m'))$.  It can be
easily checked that a $\operatorname{beta}_{m,s}(m',s')$ distribution,
for all $m'$, is unimodal with mode $m'$ if and only if $s'>0$.
Finally, we make the substitution $s = \frac{1}{\phi} - 1$ to obtain
the desired $\operatorname{beta}_{m,\phi}(\cdot,\cdot)$
parameterization, where a $\operatorname{beta}_{m,\phi}(m',\phi')$ distribution,
for all $m'$, is unimodal with mode $m'$ if and only if spread
parameter $\phi' \in (0,1)$:

\begin{equation}
\operatorname{beta}_{m,\phi}(m',\phi') =
\operatorname{beta}_{\alpha,\beta}\left(1 + \left(\frac{1}{\phi'} - 1\right)m',1 +
\left(\frac{1}{\phi'} - 1\right)\left(1-m'\right)\right).
\end{equation}

\begin{figure}[h]
\centering
\begin{subfigure}{0.48\linewidth}
\centering 
\includegraphics[width=1\linewidth]{\getFullPath{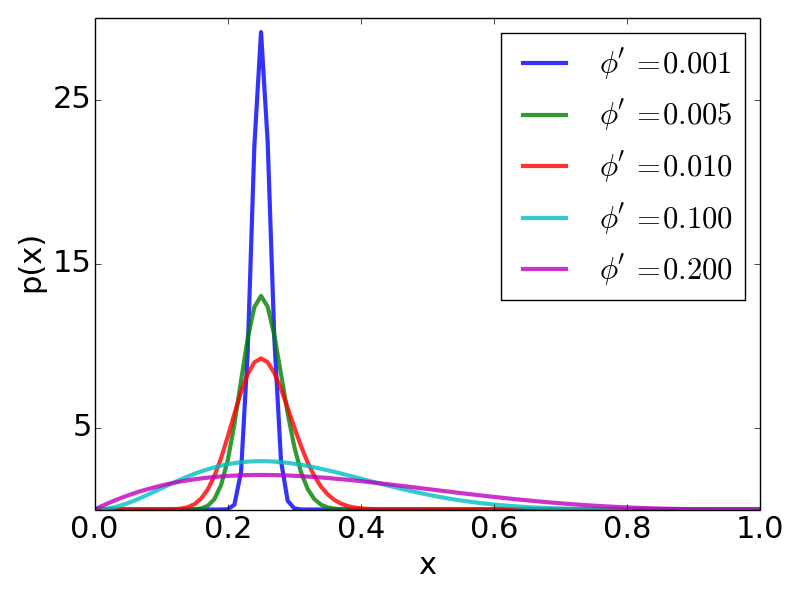}}
\caption{pdfs of $\operatorname{beta}_{m,\phi}(m',\phi')$
  distributions with $m'=0.25$}
\end{subfigure}
\quad
\begin{subfigure}{0.48\linewidth}
\centering \includegraphics[width=1\linewidth]{\getFullPath{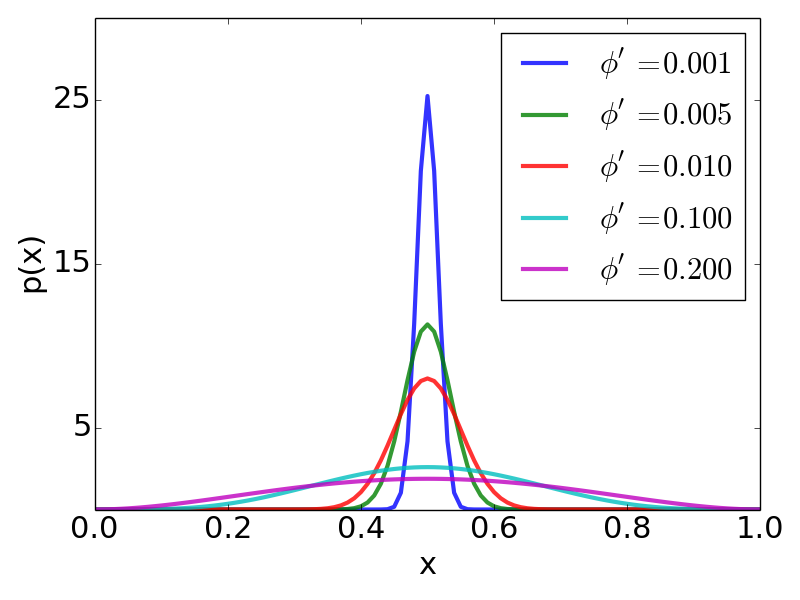}}
\caption{pdfs of $\operatorname{beta}_{m,\phi}(m',\phi')$
  distributions with $m'=0.5$}
\end{subfigure}
\caption{Illustrations of our parameterization of the beta distribution}
\label{fig:beta_dists}
\end{figure}

\begin{figure}
\centering
\begin{subfigure}{0.48\linewidth}
\centering 
\includegraphics[width=1\linewidth]{\getFullPath{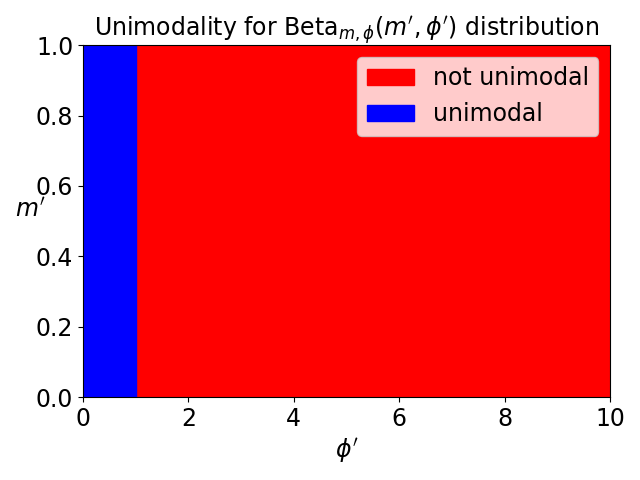}}
\caption{A $\operatorname{beta}_{m,\phi}(\cdot,\cdot)$
  parameterization suits our purposes because for a given $\phi'<1$ (a
  constraint we enforce), a
  $\operatorname{beta}_{\mu,\phi}(m',\phi')$ distribution \emph{is}
  unimodal for \emph{all} $m'$. \label{fig:good_beta}}
\label{fig:good_beta}
\end{subfigure}\hspace{1em}
\begin{subfigure}{0.48\linewidth}
\centering \includegraphics[width=1\linewidth]{\getFullPath{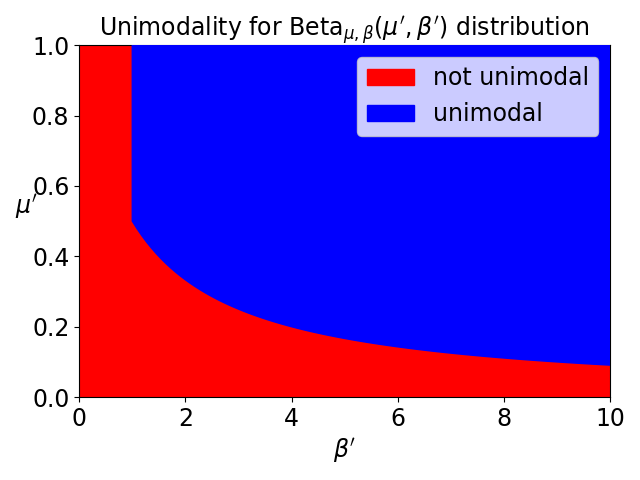}}
\caption{A $\operatorname{beta}_{\mu,\beta}(\cdot,\cdot)$ parameterization
  is unsuitable because for a given $\beta'$, there exists $m'$ such that a
  $\operatorname{beta}_{\mu,\beta}(m',\beta')$ distribution is \emph{not}
  unimodal. \label{fig:bad_beta}\\
}
\label{fig:bad_beta}
\end{subfigure}
\caption{Regions of unimodality for two different beta distribution parameterizations}
\label{fig:beta}
\end{figure}

Examples of such beta distributions are in Figure
\ref{fig:beta_dists}, and the regions of unimodality for our beta
distribution parameterization compared to the traditional one are in
Figure \ref{fig:beta}.

Our $\operatorname{gamma}_{m,\phi}(\cdot,\cdot)$ parameterization is
based on the traditional
$\operatorname{gamma}_{\alpha,\beta}(\cdot,\cdot)$ parameterization.  A
$\operatorname{gamma}_{\alpha,\beta}(\alpha',\beta')$ distribution is
unimodal with mode $\frac{\alpha'-1}{\beta'}$ if and only if
$\alpha'>1$.  Substituting $\beta = \frac{\alpha-1}{m}$ and
  $\alpha=\frac{1}{\phi}$, we obtain the
$\operatorname{gamma}_{m,\phi}(\cdot,\cdot)$ parameterization.  A
$\operatorname{gamma}_{m,\phi}(m',\phi')$ distribution is unimodal with mode $m'$
if and only if spread parameter $\phi' \in (0,1)$:

\begin{equation}
\operatorname{gamma}_{m,\phi}(m',\phi') =
\operatorname{gamma}_{\alpha,\beta}\left(\frac{1}{\phi'},
\frac{\nicefrac{1}{\phi'}-1}{m'}\right).
\end{equation}

\begin{figure}
\centering
\begin{subfigure}{0.48\linewidth}
\centering 
\includegraphics[width=1\linewidth]{\getFullPath{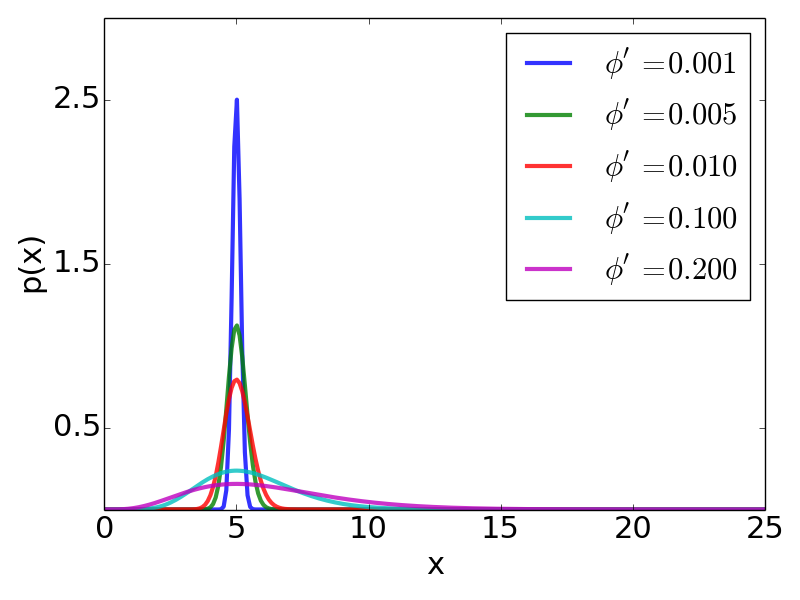}}
\caption{pdfs of $\operatorname{gamma}_{m,\phi}(m',\phi')$
  distributions with $m'=15$}
\end{subfigure}
\quad
\begin{subfigure}{0.48\linewidth}
\centering \includegraphics[width=1\linewidth]{\getFullPath{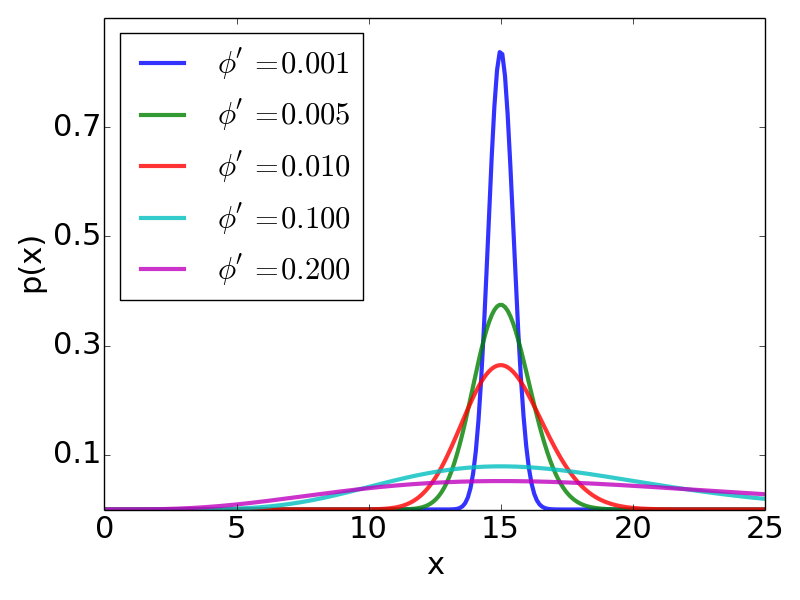}}
\caption{pdfs of $\operatorname{gamma}_{m,\phi}(m',\phi')$
  distributions with $m'=15$}
\end{subfigure}
\caption{Illustrations of our parameterization of the gamma distribution}
\label{fig:gamma_dists}
\end{figure}
Examples of such gamma distributions are in Figure \ref{fig:gamma_dists}.

Note that both a $\operatorname{beta}_{m,\phi}(m',\phi')$ and
$\operatorname{gamma}_{m,\phi}(m',\phi')$ have variance increasing in
$\phi'$.  That is why we regard $\phi'$ as a spread parameter.

\section{Additional Simulation Studies}

\begin{figure}
\centering
\begin{subfigure}{0.5\linewidth}
\centering 
\includegraphics[width=1\linewidth]{\getFullPath{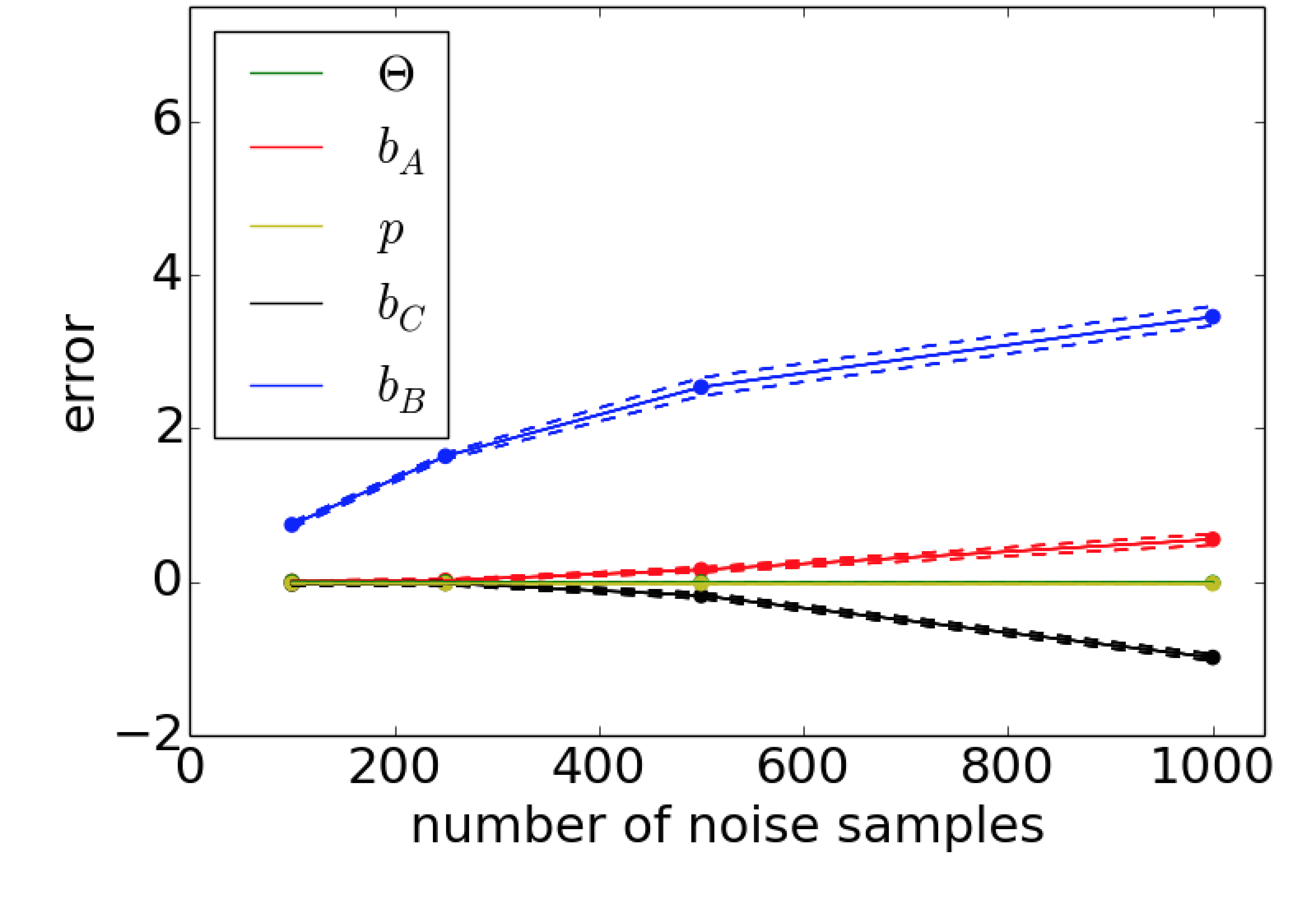}}
\end{subfigure}
\begin{subfigure}{0.5\linewidth}
\centering \includegraphics[width=1\linewidth]{\getFullPath{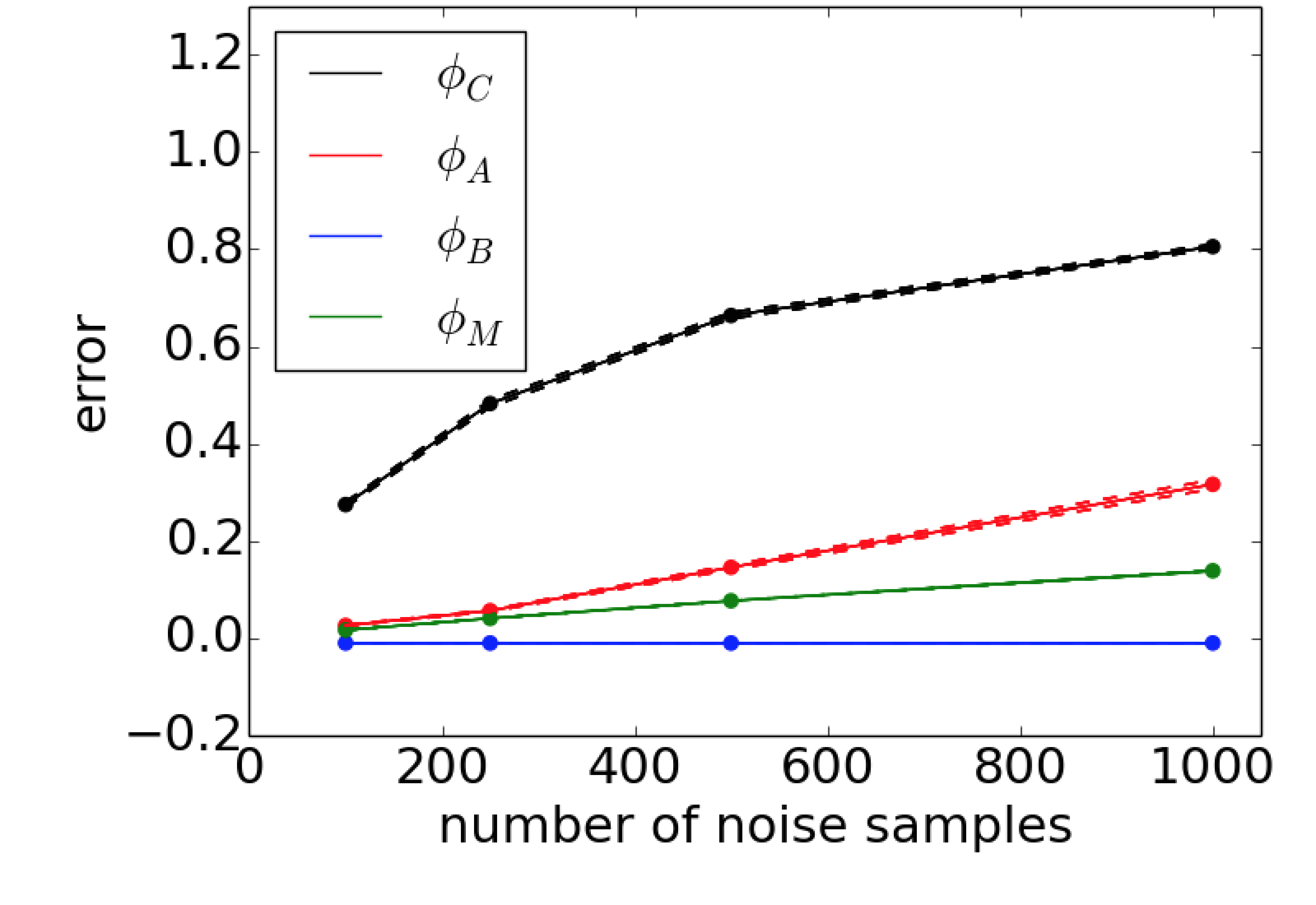}}
\end{subfigure}
\caption{For each parameter, the magnitude of the difference between posterior median parameter and true value
  increases with amount of noise data.}
\label{fig:sim_study_add_noise}
\end{figure}

Here, we examine our model's robustness to
simulated data violating the underlying model assumptions. We performed the following for various
values of $M$, the number of noise samples:  We
simulated a \emph{single} dataset by first
simulating observed function values for $5000$ samples using the same
parameters as in the first experiment.  However, then we added to the dataset an
additional $M$ \emph{noise} samples whose observed function values which were \emph{not} generated from the
model.  Instead, for those $M$ noise samples, at each time point for which
the function value is observed,
we drew the observed value uniformly from the unit interval.  For
inference, we used the same hyperparameters, sampling method, and
convergence diagnostics as before.  We then calculated, for each
parameter, the error: the \emph{signed} difference between its median posterior distribution
and the true value used to simulate (part of) the dataset.  In
Figure \ref{fig:sim_study_add_noise}, we plot for each parameter this
signed difference (as well as the 25-th and 75-th percentile of the error)
as $M$, the number of noise data samples, varies.  As expected, the
magnitude of this error is
increasing in $M$.

\begin{figure}
\centering
\begin{subfigure}{0.5\linewidth}
\centering 
\includegraphics[width=1\linewidth]{\getFullPath{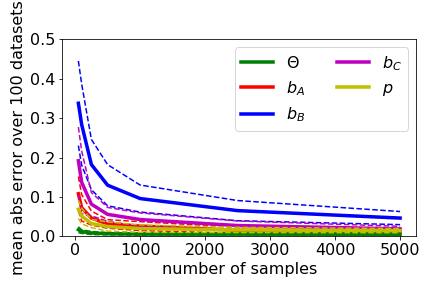}}
\end{subfigure}
\begin{subfigure}{0.5\linewidth}
\centering \includegraphics[width=1\linewidth]{\getFullPath{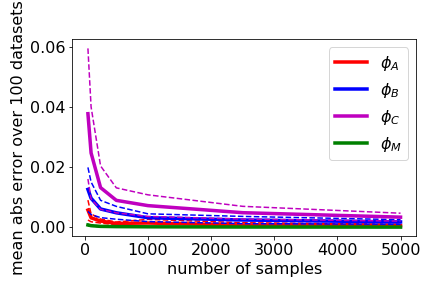}}
\end{subfigure}
\caption{For each parameter, the mean absolute error over the
  simulated datasets decreases with the size of the simulated
  datasets.  Dotted lines denote 1 standard deviation.}
\label{fig:sim_study_vary_N_unsigned}
\end{figure}

We also include here an additional plot describing the results of the
simulation study in Section \ref{sec:sim} of the manuscript.  Recall
that the purpose of the simulation study was to examine
the ability of our model to recover the model parameters as the amount of
data simulated using those parameters grew. We chose a
single set of shared model parameters and hyperparameters
$\mu_A,\mu_B,\mu_C$.  Then, we performed the following for several
values of $N$, the number of patients in a simulated dataset:  We simulated
100 datasets, where for each dataset we used that set of chosen parameters to simulate observed function values
$y_i(t)$ for $N$ patients at times $t \in
\{1,2,4,8,12,18,24,30,36,42,48\}$,  the same times at which data were observed in the prostate cancer
dataset.  For each dataset, we obtained for each parameter a point
prediction as its posterior median, and calculated the \emph{unsigned} error.  Figure \ref{fig:sim_study_vary_N_unsigned} shows the mean and
standard deviation of the signed error across the 100 datasets for each
parameter, for various values of N (N$\in\{50,100,250,500,1000,2500,5000\}$).  Note that in the manuscript, we
had shown the analogous results for the \emph{signed} error.  Please
see the manuscript for further details on the values of the parameters
used to simulate the data as well as the inference method.

\section{Dataset Filtering\label{sec:filtering}}

We apply two filters to our dataset such that the
retained patients possess the properties required of the target population, retaining 237 out of the original 304 patients' data.  We note that
while our target population includes patients merely considering a
prostatectomy, our actual dataset only contains data from patients who actually
underwent a prostatectomy.  Our assumption is that patients passing
the two filters will belong to the target population, not differing
from it on any unmeasured covariates impacting sexual function
trajectory.  For the first filter, we removed 17 patients whose pre-treatment sexual function
level was less than 0.1, as our target population was defined to have
a non-negligible pre-treatment function level.  Secondly, as our
dataset contains patients who both did and did not use an erectile aid
post-treatment, we wish to remove patients who did use an erectile
aid.  Unfortunately, our dataset did not record whether an aid was
used, and we had to instead infer its usage from the data, based on a
criteria set by the urologist co-author: For the second filter, we removed patients whose representative curve, obtained by fitting a curve
(under least squares error, where the asymptotic level was not constrained) to their data points, was, at 48 months post-treatment (the last
time for which data was available), higher than their pre-treatment
value.  Doing so removed an additional 50 patients, so that our final
retained dataset contained 237 out of the original 304 patients.
This second filter is justified by the urologist co-author's past clinical
experience that a prostatectomy patient who does not use an erectile
aid has a negligible chance of having their long-term sexual function
exceed that pre-treatment.  Thus, this filter will remove a neglible
number of patients that did not use an erectile aid, though actually
it retains some patients not in the target population - those who used
an aid, but whose sexual function did not improve long-term.

\section{Exploratory Plots}
\begin{figure}
\centering
\begin{subfigure}{0.5\linewidth}
\centering 
\includegraphics[width=1\linewidth]{\getFullPath{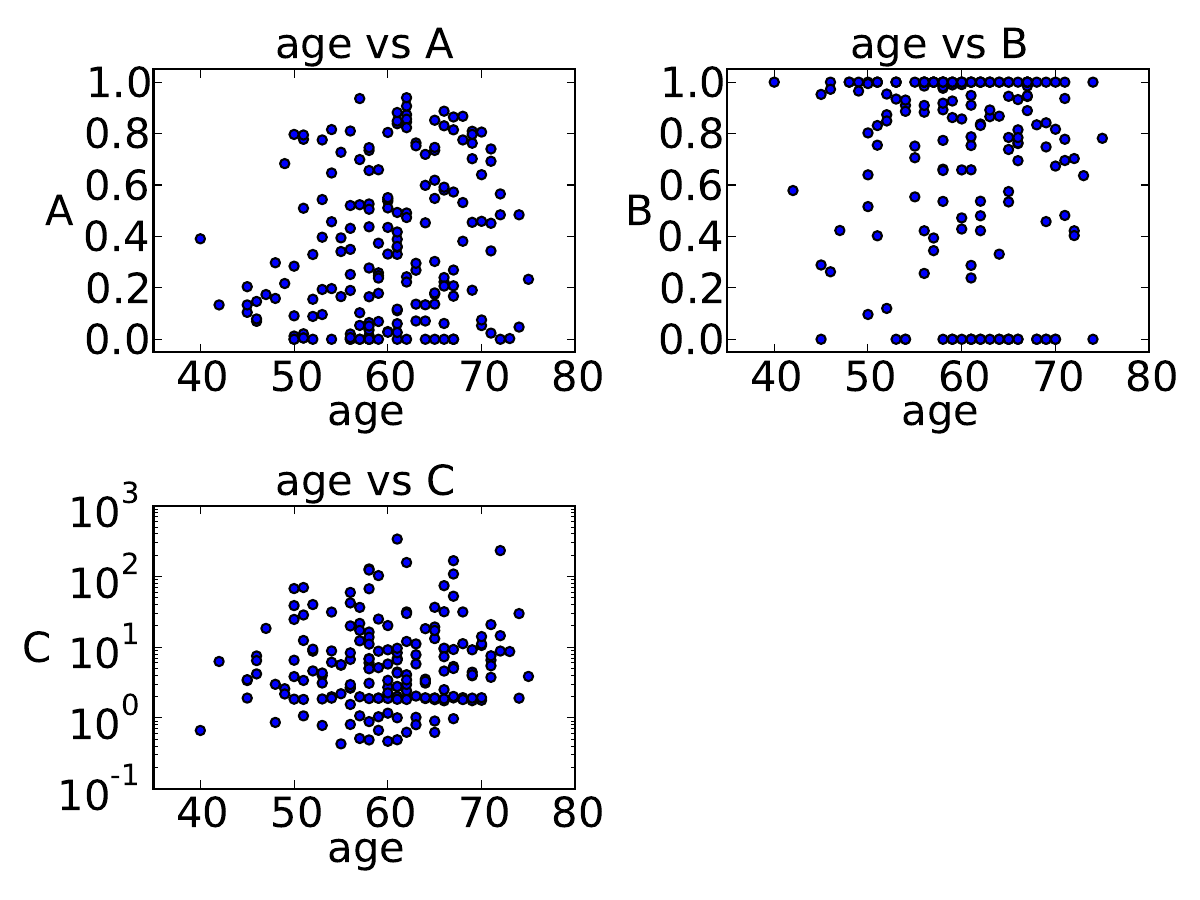}}
\caption{parameters vs age}
\label{fig:age_scatter}
\end{subfigure}
\begin{subfigure}{0.5\linewidth}
\centering \includegraphics[width=1\linewidth]{\getFullPath{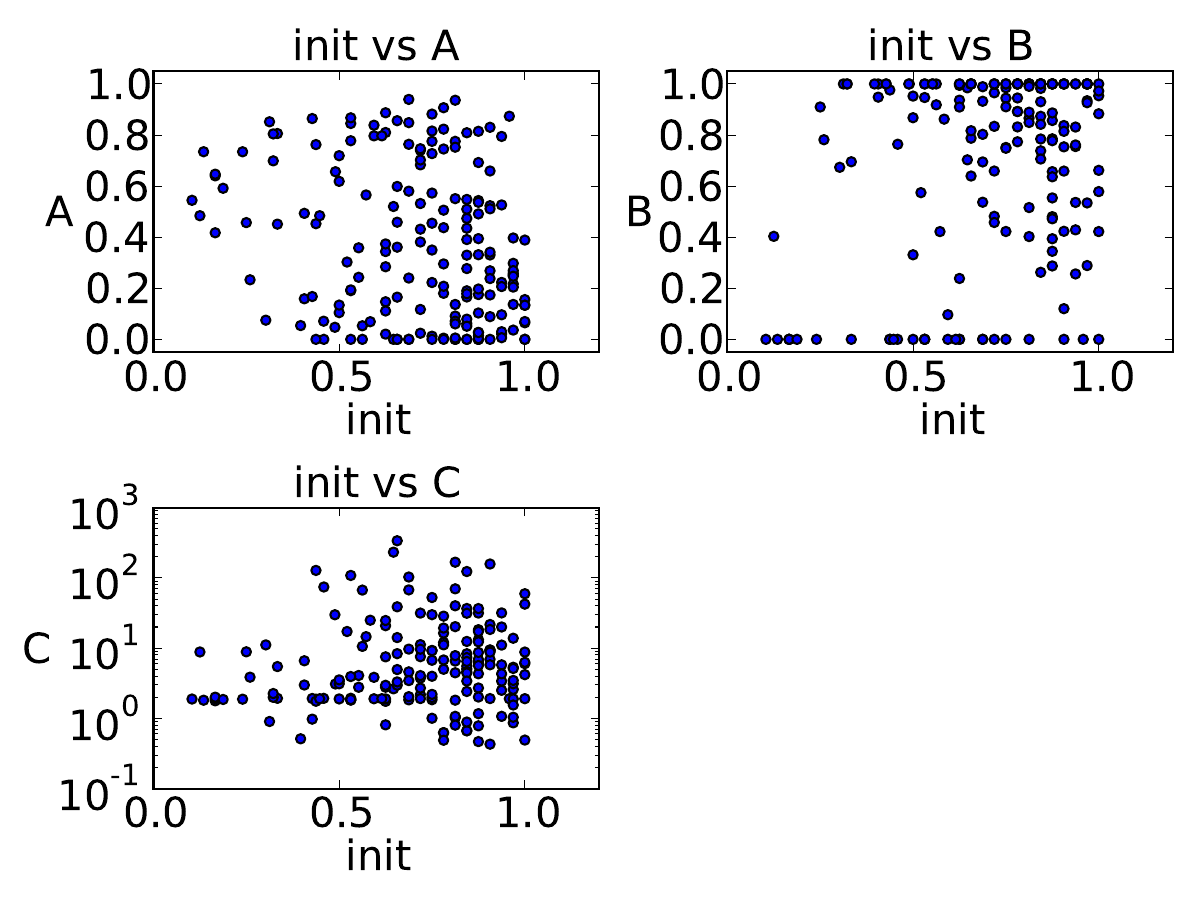}}
\caption{parameters vs pre-treatment level}
\label{fig:init_scatter}
\end{subfigure}
\caption{The three curve parameters show some dependence on a patient's age and pre-treatment sexual function value.}
\label{fig:scatters}
\end{figure}

\begin{figure}
\centering
\begin{subfigure}{0.48\linewidth}
\centering 
\includegraphics[width=1\linewidth]{\getFullPath{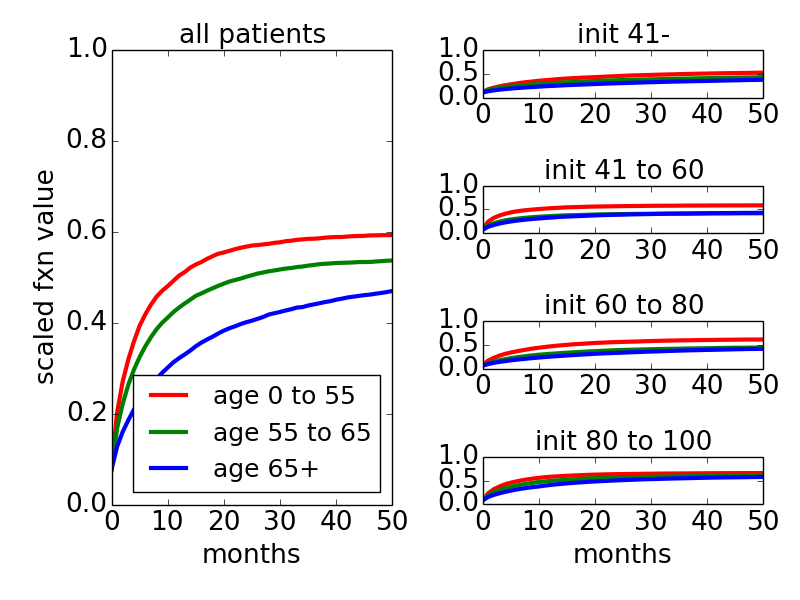}}
\caption{Averaged scaled function values stratified by age}
\end{subfigure}
\quad
\begin{subfigure}{0.48\linewidth}
\centering \includegraphics[width=1\linewidth]{\getFullPath{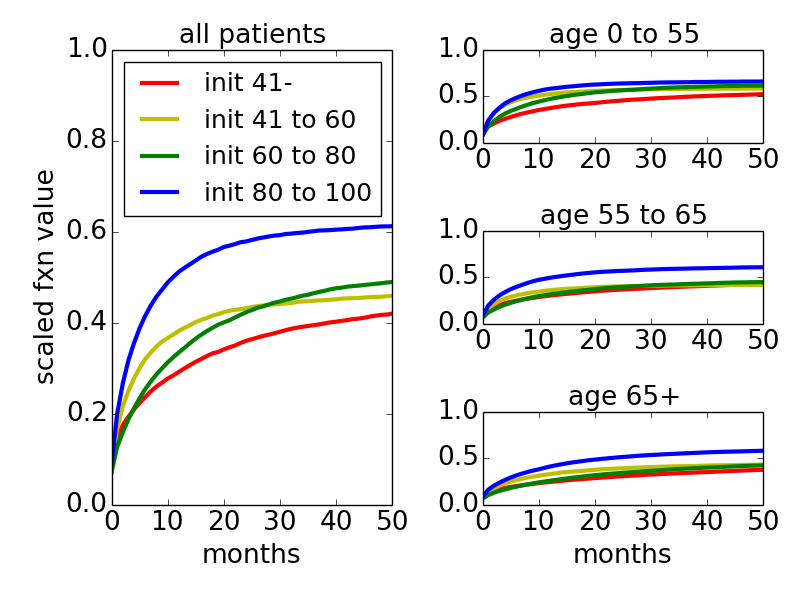}}
\caption{Averaged scaled function values stratified by pre-treatment value}
\end{subfigure}
\caption{Plotting the average scaled function values over the dataset stratified by age and pre-treatment sexual function value suggests the dependence of latent recovery shapes on age and pre-treatment value.}
\label{fig:stratified_real_data}
\end{figure}

To identify potential correlates of recovery curve shapes, for every
patient, we used curve fitting
to find the $A,B,C$ parameters corresponding to their post-event
recovery shapes.  We made scatter plots of each of those parameters
against all available covariates to identify ones that correlated with
curve parameters, and identified the pre-treatment sexual function
level (referred to as ``init'' in all figures) and patient age (at
treatment time) to be the 2 covariates most strongly correlated with
curve parameters.  The scatter plots relating curve parameters and
those 2 covariates are in Figure \ref{fig:scatters}, and guided the
creation of binned categorical features based on patient age and
pre-treatment function level.  
To visualize the effect of these 2
covariates on recovery shape from another view, we stratified the
patients by age bin and pre-treatment sexual function level
bin, and plotted in Figure \ref{fig:stratified_real_data} the average shape of the patients in each
bin.  (The ranges of the of the bins we used in the categorical
features are contained in the figure.)

\section{Modeling Scaled Function Values\label{sec:scaled}}
In general there are 2 potential approaches to modeling patients'
absolute function values - we can do so directly, or instead model
their \emph{scaled} function value (absolute function value divided by
pre-treatment function value), and then scale it by their
pre-treatment value.  As we take the latter approach, we examine the
prostatectomy dataset to justify doing so, showing that naive models
that model scaled function value have superior in-sample performance
on the prostatectomy dataset to their unscaled analogue.  The measure of performance we consider here and throughout the rest of the paper is loss as given by average absolute prediction error.

\begin{figure}
\centering
\begin{subfigure}{0.5\linewidth}
\centering 
\includegraphics[width=1\linewidth]{\getFullPath{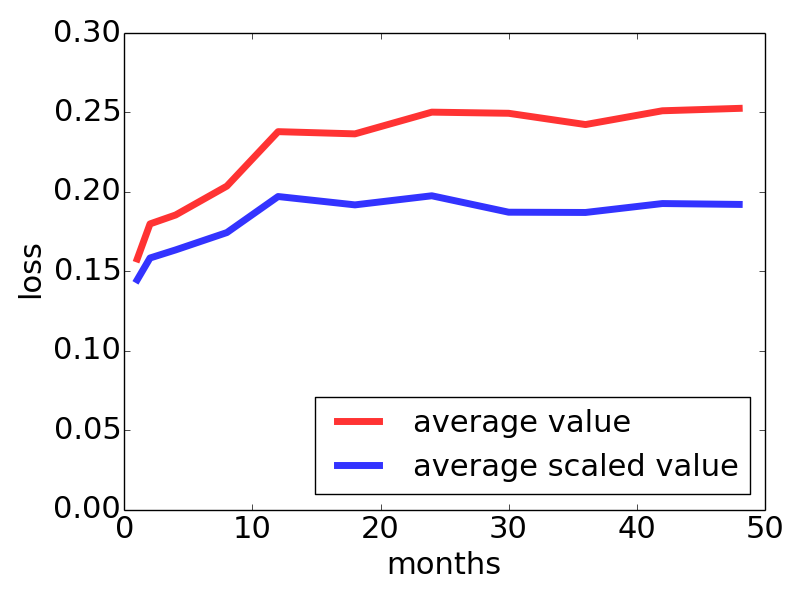}}
\caption{covariate-independent models}
\label{fig:scaled_prior}
\end{subfigure}
\begin{subfigure}{0.5\linewidth}
\centering \includegraphics[width=1\linewidth]{\getFullPath{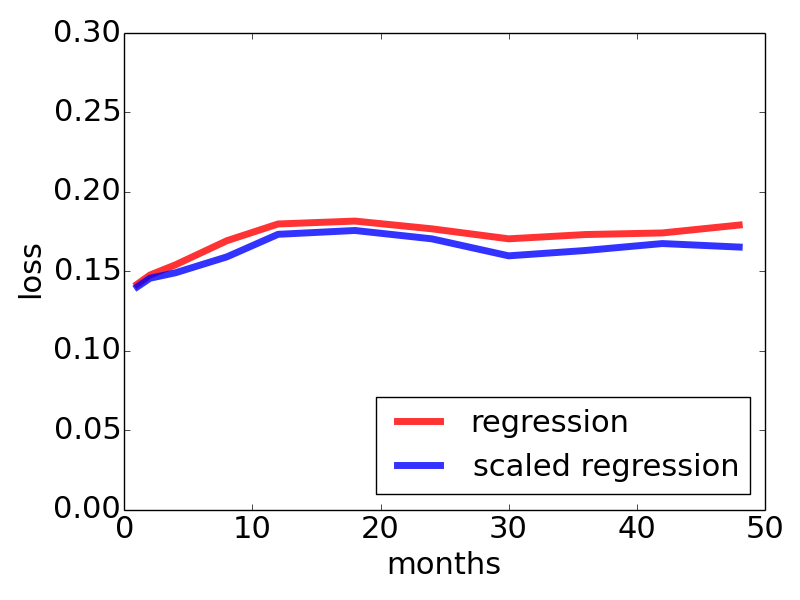}}
\caption{regression models}
\label{fig:scaled_regression}
\end{subfigure}
\caption{In-sample performance is better when using models that model scaled function value as opposed to absolute function value.}
\label{fig:abs_vs_scaled}
\end{figure}

First, we consider a baseline model where at each timepoint, each patient is
predicted to possess the average absolute function value (over the
dataset) at that timepoint (labeled ``average value'').  We compare
its in-sample predictive performance (fit) to a model where each patient is predicted to have the average \emph{scaled} function value at that timepoint (labeled ``average scaled value'') in Figure \ref{fig:scaled_prior}.  Next, we plot in Figure \ref{fig:scaled_regression} the in-sample predictive performance of two analogous models that use separate generalized linear regression models at each of the 11 common timepoints to model a patient's absolute and scaled function value, respectively, as a function of patient features.  In other words, one model (labeled ``regression'') regresses the absolute function value at each timepoint against patient features, and the other model (labeled ``scaled regression'') regresses the scaled function value against features, and to obtain a prediction of absolute function value, assumes that a patient's absolute function value is their scaled function value divided by their pre-treatment function value.  The generalized linear regression models employ a logistic inverse link function and assume a normal response distribution.  As can be seen, in both the cases, the models that predict scaled function value have superior predictive performance.

\section{Additional Analysis of Prostatectomy Dataset}
In the following subsections, we perform additional analysis of the
prostatectomy dataset as it relates to our model.

\subsection{Sensitivity of Out-Of-Sample Performance to
  Hyperparameters\label{sec:hyper_performance}}

\begin{figure}
\centering
\begin{subfigure}{0.48\linewidth}
\centering 
\includegraphics[width=1\linewidth]{\getFullPath{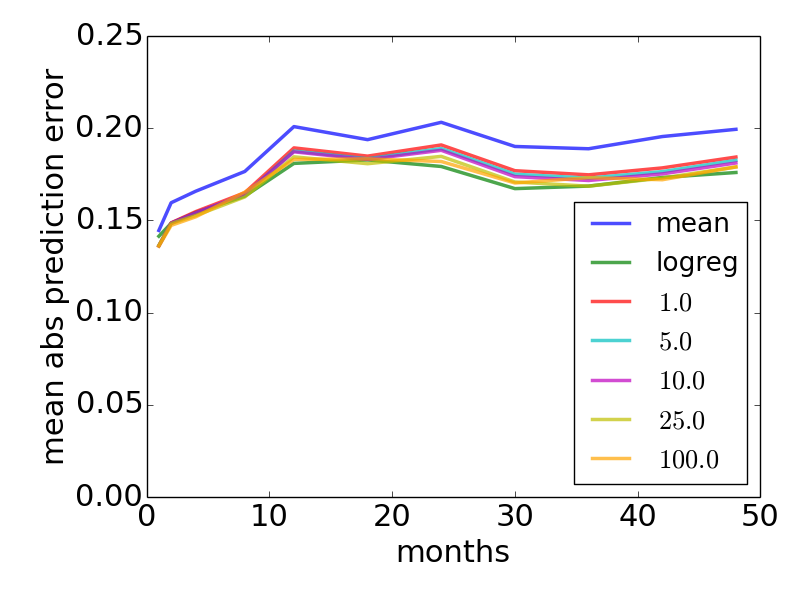}}
\caption{Out-of-sample performance as the tied values of
  $l_A,l_B,l_C$ vary.  The numerical labels indicate the tied value of $l_A,l_B,l_C$.}
\label{fig:sensitivity_phi}
\end{subfigure}
\quad
\begin{subfigure}{0.48\linewidth}
\centering \includegraphics[width=1\linewidth]{\getFullPath{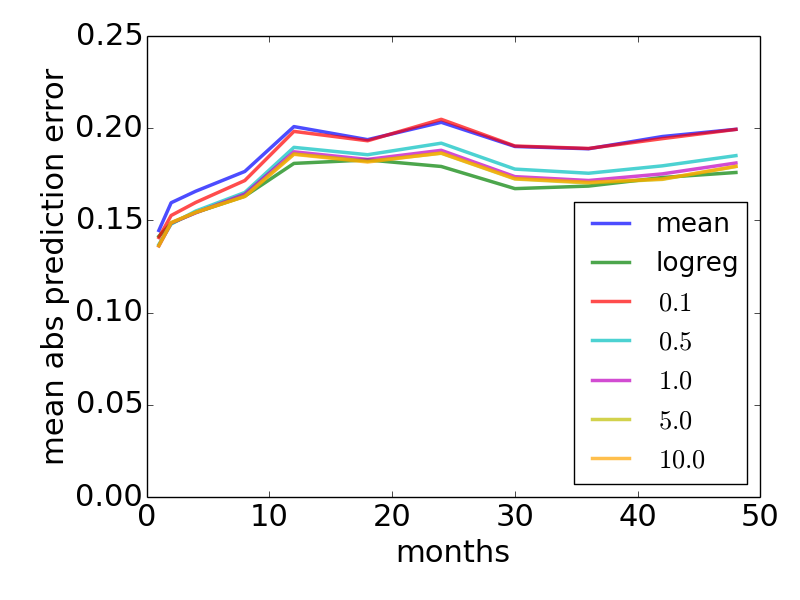}}
\caption{Out-of-sample performance as the tied values of $s_A,s_B,s_C$
  vary.  The numerical labels indicate the tied value of $s_A,s_B,s_C$.}
\label{fig:sensitivity_sigma}
\end{subfigure}
\caption{Sensitivity of out-of-sample performance to hyperparameters.}
\label{fig:sensitivity}
\end{figure}

We provide a sense in Figures \ref{fig:sensitivity_phi} and
\ref{fig:sensitivity_sigma} how out-of-sample performance changes with
the values of hyperparameters $s_A,s_B,s_C$ and $l_A,l_B,l_C$.  The joint setting of
those hyperparameter values we used in the analysis of the
prostatectomy data was
$l_A=l_B=l_C=10$, 
$s_A=s_B=s_C=1.0$ (and $l_M=10$, though performance was not sensitive
to $l_m$).  In Figure \ref{fig:sensitivity_phi},
we tie $\l_A=\l_B=l_C$, and
illustrate how out-of-sample performance changes as the tied
value of $l_A,l_B,l_C$ changes, with the remaining hyperparameters set
to their original values.  In Figure
\ref{fig:sensitivity_sigma}, we tie $s_A=s_B=s_C$ and illustrate how
out-of-sample performance changes as the tied
value of $s_A,s_B,s_C$ changes, with the remaining hyperparameters set
to their original values.  In both figures, we also include for
comparison the
performance of the ``scaled mean'' (denoted by ``mean'' in the figure)
and ``scaled regression'' (denoted by ``logreg'' in the figure) models as
described in Section 6.4 of the main text.

\subsection{Sensitivity of posterior predictive variance to
  hyperparameters}
\begin{figure}
\centering
\includegraphics[width=0.48\linewidth]{\getFullPath{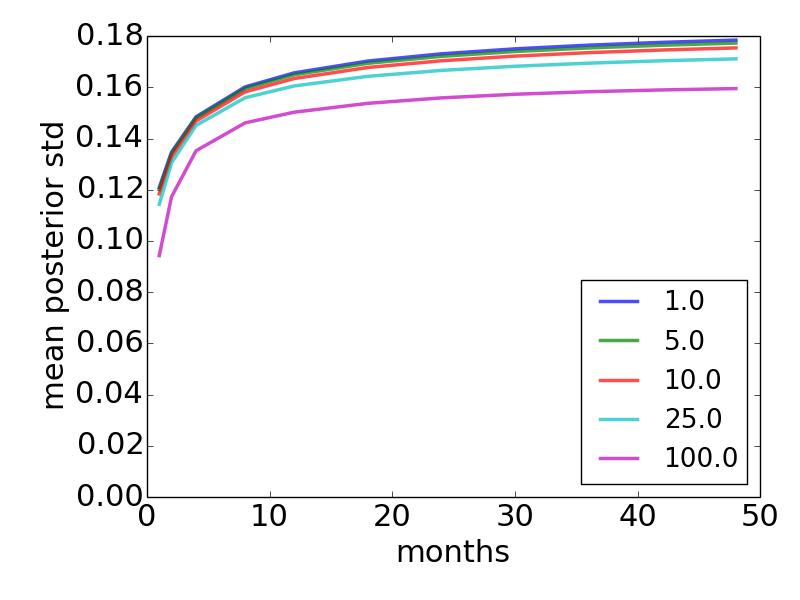}}
\caption{The standard deviation of the posterior predictive distribution, averaged over the
  dataset, is robust to the tied value of $l_A,l_B,l_C$, indicated by
  the labels.}
\label{fig:variance}
\end{figure}

Here, we examine the sensitivity of posterior predictive variance to
variance hyperparameters $l_A,l_B,l_C$. In particular,
we fix $s_A=s_B=s_C=1.0$ and $l_m=10$.  Then, for several values of
$v$, we set $l_A=l_B=l_C=v$, and for each time point $t$ for which there
was recorded data, calculated the following: for each patient $i$ in
the dataset, we calculated the
standard deviation of the posterior predictive distribution of their ``underlying''
function value at time $t$, $f_i(t)$ , and then averaged these standard
deviations over the entire dataset to get the mean standard
deviation of the posterior predictive distribution at time $t$.  In Figure \ref{fig:variance}, we plot, for
several values of $v$ (the legend indicates $v$), the mean posterior
standard deviation over time.

\subsection{Posterior Predictive Checks\label{sec:checks}}

To
assess model fit, we performed standard posterior predictive checks with
the following test statistics of the data:
\begin{itemize}[noitemsep]
  \item average total drop: pre-treatment level minus level at the last
    available time, averaged over patients.
  \item average relative total drop: pre-treatment level minus level at the last
    available time, divided by pre-treatment level,
    averaged over patients.
  \item average initial drop: pre-treatment level minus level at the first
    available time, averaged over patients.
  \item average relative initial drop: pre-treatment level minus level at the first
    available time, divided by pre-treatment level, averaged over patients.
  \item average recovery: level at last available time minus level at first
    available time post-treatment, averaged over patients
  \item average relative recovery: level at last available time minus level at first
    available time post-treatment, divided by pre-treatment level, averaged over patients
\end{itemize}

\begin{figure}
\centering
\includegraphics[width=1.0\linewidth]{\getFullPath{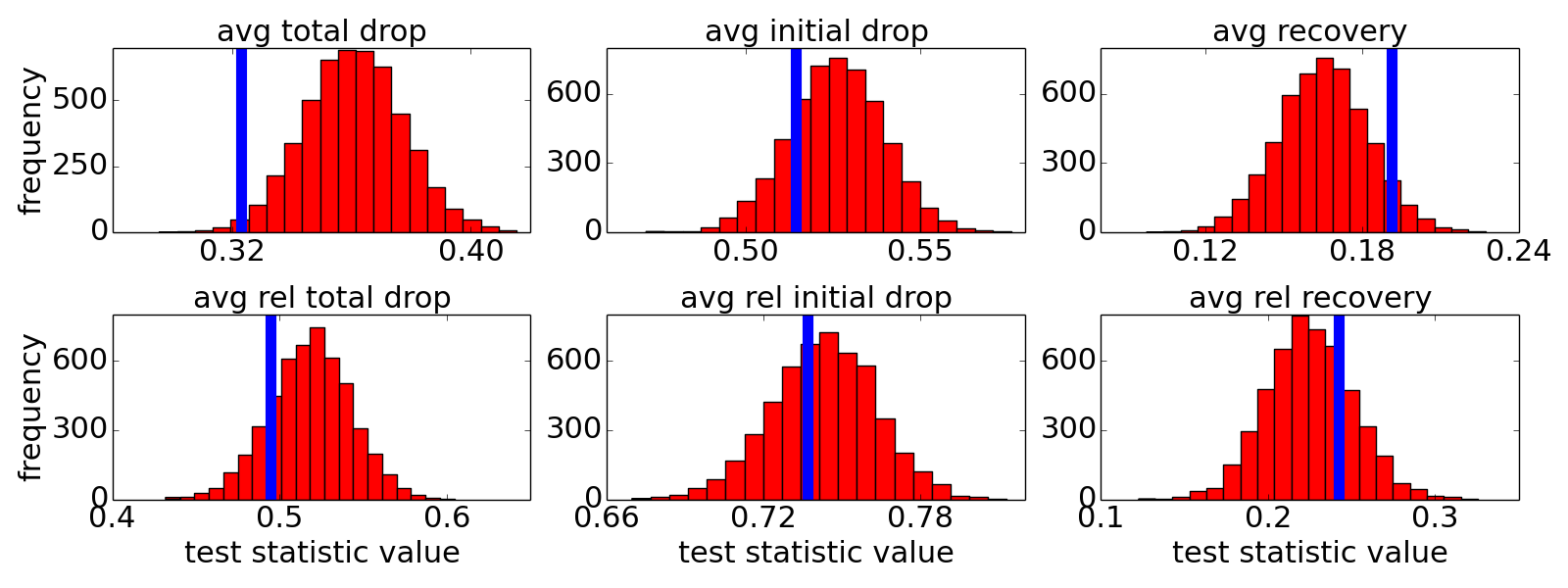}}
\caption{The empirical ``relative'' test statistics (blue vertical lines) are not
  extreme relative to that of the posterior over replicate datasets.}
\label{fig:agg_model_fit}
\end{figure}

In Figure \ref{fig:agg_model_fit}, the distribution of the test statistic over replicate datasets is
shown in red for each test statistic, as is the actual test statistic
for the data in blue.  We see that under the 3 test statistics that are
defined \emph{relative} to the pre-treatment value, the data are fit
quite well.  Though, this is less the case for the other 3 test statistics that are
not defined relative to the pre-treatment value.  However, we believe
the pattern of recovery relative to the pre-treatment value is a more
important feature of the data than the absolute pattern of recovery -
this is in part why we chose to model the relative function
level rather than the absolute function level.

\begin{figure}
\centering
\includegraphics[width=1.0\linewidth]{\getFullPath{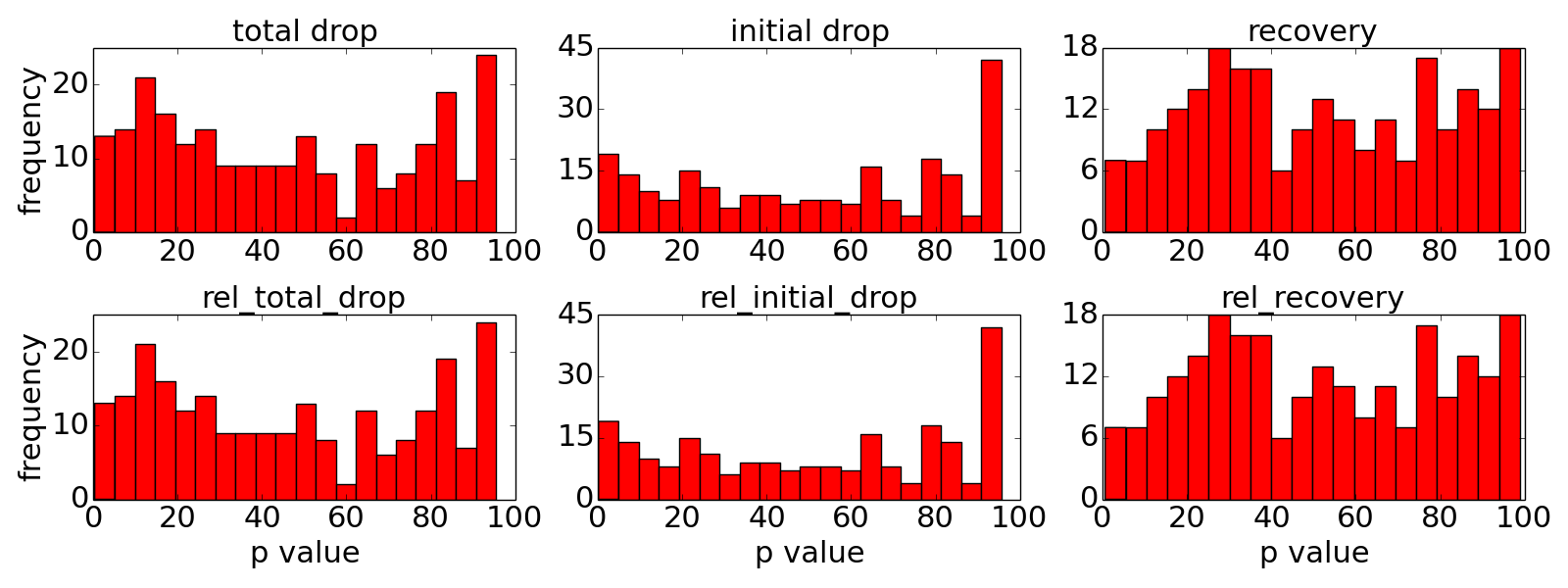}}
\caption{Most patients do not have an extreme empirical
  patient-specific test statistic relative to that of the posterior
  over replicate datasets, aside from the initial (relative) drop test
  statistic.}
\label{fig:ind_model_fit}
\end{figure}

The above analysis addresses how the dataset as a whole is fit by the
model.  To assess how well individual patients' data are fit by the
model, we then defined a test statistic that is a function of only a
single patient's data.  We then calculated the test statistic for each
patient, thus obtaining 237 patient-specific test statistic values.  Then, for each of
the 237 patients, we calculated the p-value of
their actual test statistic relative to that under replicate datasets
simulated from the posterior.  We then report in Figure
\ref{fig:ind_model_fit} the distribution of the p-values, for that test
statistic, over the patients in the dataset.  These test statistics
are the similar to before, except without averaging (they are all
functions of a single patient's data):
\begin{itemize}[noitemsep]
  \item total drop: pre-treatment level minus level at the last
    available time
  \item initial drop: pre-treatment level minus level at the first
    available time
  \item recovery: level at last available time minus level at first
    available time post-treatment
  \item relative total drop: pre-treatment level minus level at the last
    available time, divided by pre-treatment level.
  \item relative initial drop: pre-treatment level minus level at the first
    available time, divided by pre-treatment level.
  \item relative recovery: level at last available time minus level at first
    available time post-treatment, divided by pre-treatment level.
\end{itemize}

\begin{figure}[h]
\centering
\begin{subfigure}{0.5\linewidth}
\centering 
\includegraphics[width=1\linewidth]{\getFullPath{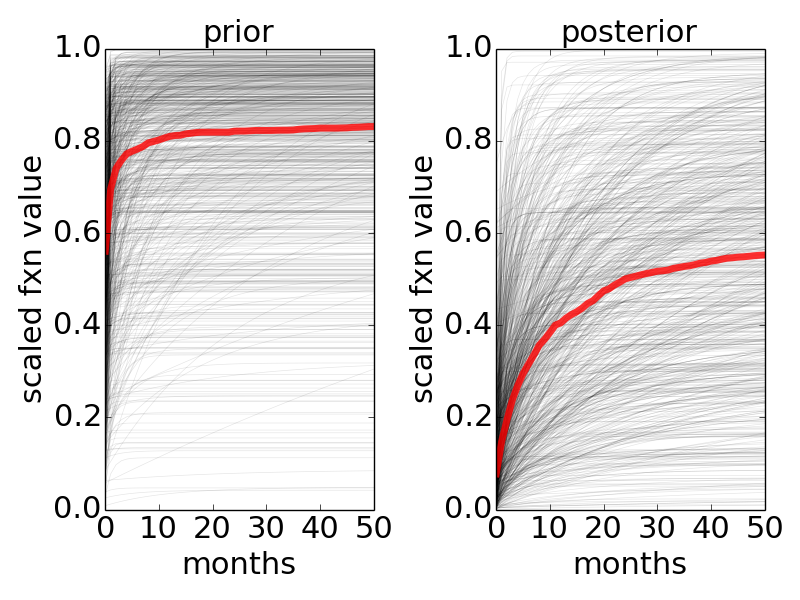}}
\end{subfigure}
\begin{subfigure}{0.5\linewidth}
\centering \includegraphics[width=1\linewidth]{\getFullPath{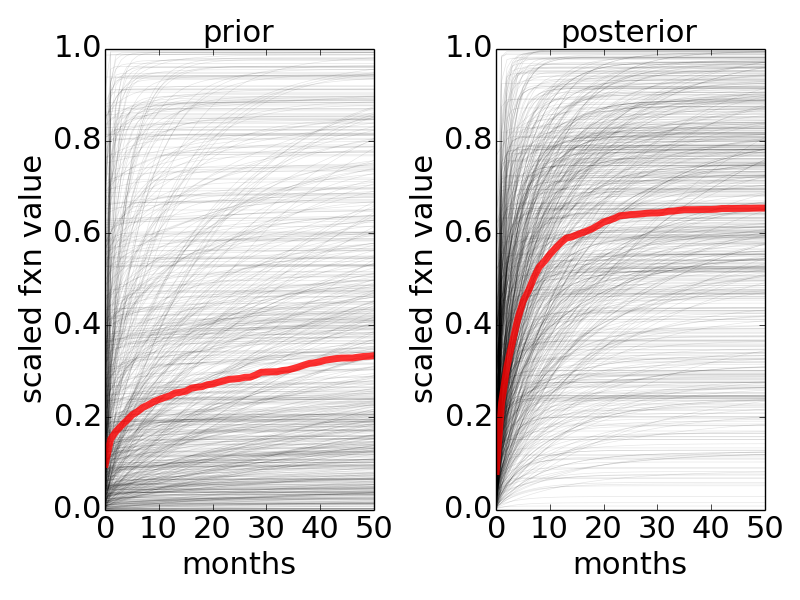}}
\end{subfigure}
\caption{The prior vs posterior predictive distribution for 2 patients.}
\label{fig:prior_vs_posterior}
\end{figure}

\begin{figure}[h]
\centering
\includegraphics[width=0.48\linewidth]{\getFullPath{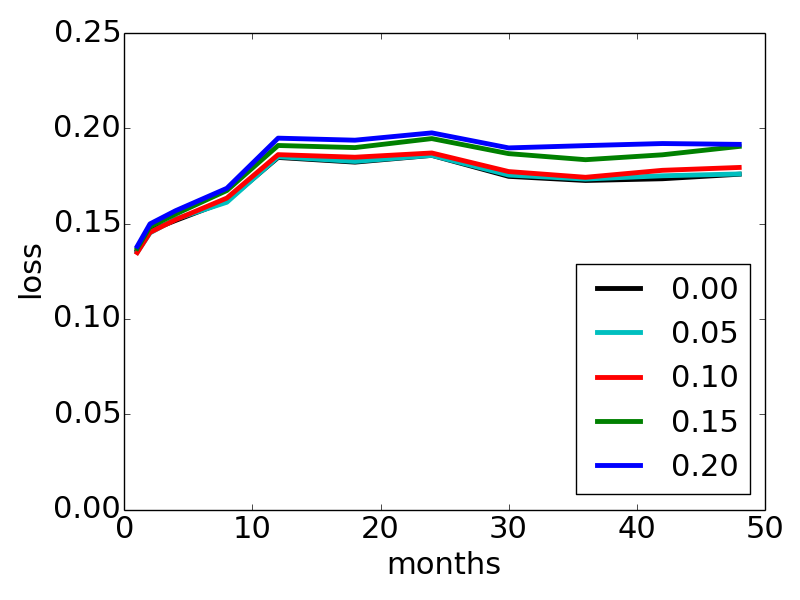}}
\caption{Predictive performance is robust to moderate measurement
  error in pre-treatment function level.  Labels indicate the standard
deviation of the perturbation noise distribution.}
\label{fig:jitter_s}
\end{figure}

We see that for most patients, their (relative) total drop and
(relative) recovery are not
too extreme relative to the respective values in the replicate
datasets.  However, for for roughly
45 patients, their (relative) initial drop was at the 95th percentile of the
respective initial drops in the replicate datasets, indicating that
for those patients, their initial drop was greater than that predicted
by the fit model.  In fact, every single one of those 45 patients
reported a sexual function value of exactly 0 at the first time for
which data was available post-treatment.  The likelihood allows the
reported values, $y^{(i)}(t)$, to take on a value 0, with a probability that is
constant across patients and time.  One remedy would be to allow this
probability to depend on patients and time, perhaps through the
patient's underlying function value, $f^{(i)}(t)$.  However, as this
feature of the data is likely noise, we opted not to do so.

\subsection{Impact of the Data on Posterior}
To understand the impact of
the data on the posterior predictivedistribution, we illustrate for two
patients their prior and posterior distribution over
recovery curves in Figure \ref{fig:prior_vs_posterior}.

\subsection{Robustness to Measurement Error of Pre-treatment Function
  Level}

In our model, we assume the pre-treatment function level is known.
Here, we study the effect of measurement error in
the pre-treatment function level on predictive performance.  In
particular,
we performed an experiment where for each patient $i$ in our dataset,
we obtained a draw from a normal distribution with mean 0 and fixed
standard deviation (denoted $c$), and perturbed $S_i$ by that random
amount (by choosing a
perturbation direction of up or down randomly, then moving in that
direction, reflecting off the boundaries of 0 and 1 if necessary) to obtain a single perturbed dataset.
For a fixed value of $c$, we then obtained 8 perturbed datasets, and
can then plot the cross validation performance over time (using the same
evaluation metric as in evaluating out-of-sample performance in
Section 6.4 of the main text), averaged over the
8 datasets.  Figure \ref{fig:jitter_s} then shows the average cross
validation performance, for several values of $c$.

\clearpage

\subsection{Posterior Predictive Distributions for Patients}
In the following pages, we display the posterior predictive
distributions for all 12 types of patients in the prostatectomy
dataset.  Recall that patients belong to 1 of 3 strata based on their
age, and 1 of 4 strata based on their pre-treatment level (see Figure
12 of the manuscript), which gives a total of 12 types of patients.  The titles
of the plots indicate the age and pre-treatment level strata of the
patient whose posterior predictive distribution is displayed (the
notation is the same as in Figure 13 of the manuscript.)

\clearpage
\includepdf[pages=-]{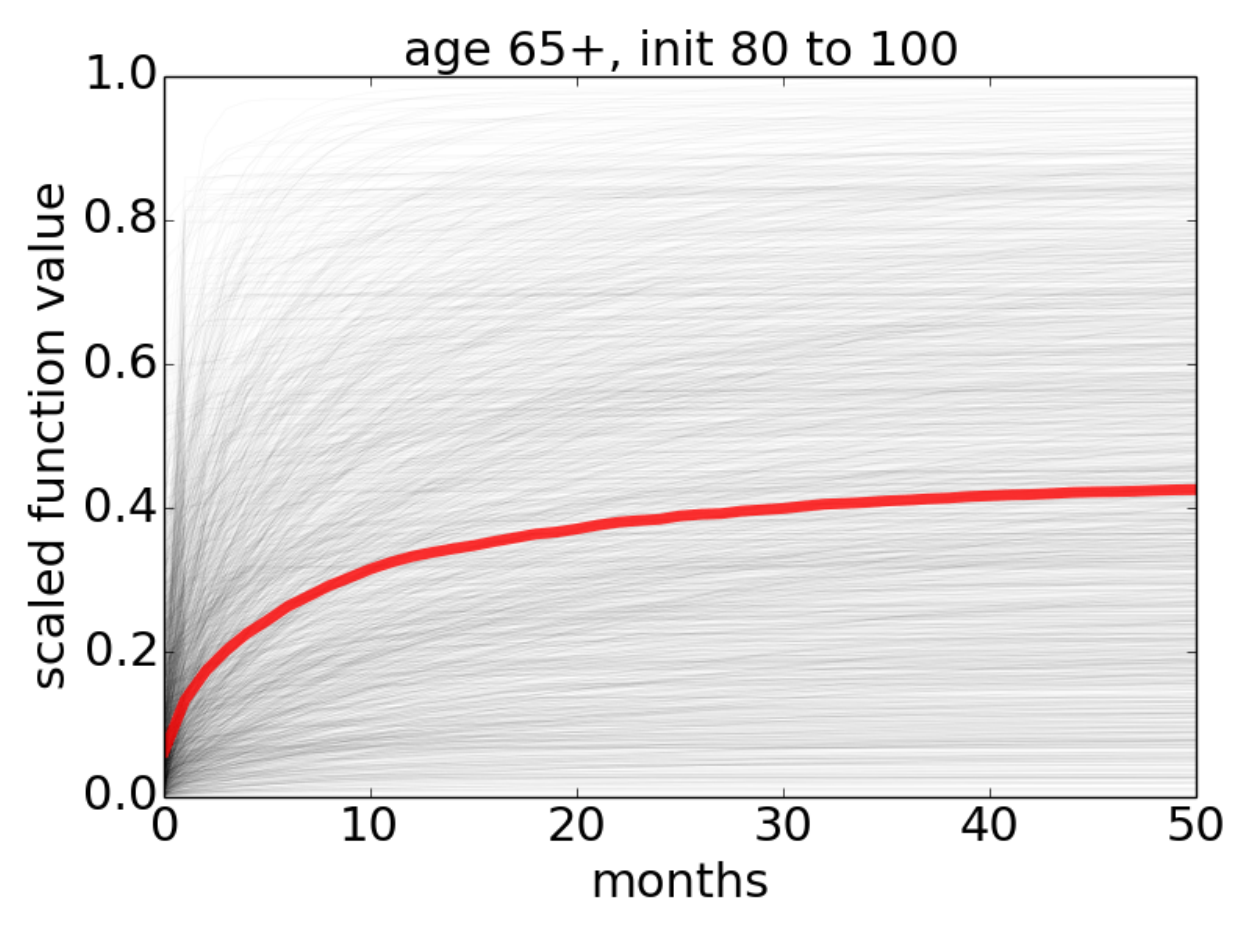}

\end{document}